\documentclass[fleqn,usenatbib]{mnras}

\usepackage{newtxtext,newtxmath}
\usepackage[T1]{fontenc}
\usepackage{hyperref}
\usepackage{subcaption}
\usepackage{caption}
\usepackage{orcidlink}
\usepackage{acronym}
\usepackage{url}
\acrodef{3g}[3G]{third-generation}

\acrodef{bh}[BH]{black hole}
\acrodef{bbh}[BBH]{binary black hole}
\acrodef{cbc}[CBC]{compact binary coalescence}
\acrodef{cdm}[CDM]{cold dark matter}
\acrodef{cmb}[CMB]{cosmic microwave background}
\acrodef{de}[DE]{dark energy}
\acrodef{DEMNUni}[DEMNUni]{Dark Energy and Massive Neutrino Universe}
\acrodef{dm}[DM]{dark matter}
\acrodef{em}[EM]{electromagnetic}
\acrodef{eos}[EoS]{equation of state}
\acrodef{et}[ET]{Einstein Telescope}
\acrodef{fof}[FoF]{friend-of-friends}
\acrodef{gw}[GW]{gravitational wave}
\acrodef{hbi}[HBI]{hierarchical Bayesian inference}
\acrodef{hi}[\textrm{H\textsc{i}}]{neutral hydrogen}
\acrodef{im}[IM]{intensity mapping}
\acrodef{los}[LoS]{line-of-sight}
\acrodef{lss}[LSS]{large-scale structure}
\acrodef{pe}[PE]{parameter estimation}
\acrodef{skao}[SKAO]{SKA Observatory}
\acrodef{sne}[SNe]{supernovae}
\acrodef{snr}[SNR]{signal-to-noise ratio}

\definecolor{SKABlue}{RGB}{24, 0, 104}
\hypersetup{colorlinks=true,                citecolor=SKABlue,
            linkcolor=SKABlue, urlcolor=SKABlue, linktocpage=true}
\newcommand{\dd}{{\rm d}}
\newcommand{\hi}{\textrm{H\textsc{i}}}

\DeclareRobustCommand{\VAN}[3]{#2}
\let\VANthebibliography\thebibliography
\def\thebibliography{\DeclareRobustCommand{\VAN}[3]{##3}\VANthebibliography}

\usepackage{graphicx}	% Including figure files
\usepackage{xcolor}
\usepackage{amsmath}	% Advanced maths commands
\usepackage{booktabs}
\usepackage{tabularx}
\usepackage{makecell}
\usepackage[table]{xcolor}
\title[Radio sirens]{Radio sirens: inferring $H_0$ with binary black holes and neutral hydrogen in the era of the Einstein Telescope and the SKA Observatory}

\author[Dupletsa et al.]{Ulyana Dupletsa$^{1}$\thanks{E-mail: ulyana.dupletsa@oeaw.ac.at}\orcidlink{0000-0003-2766-247X},
Simone Mastrogiovanni$^{2}$\orcidlink{0000-0003-1606-4183},
Marta Spinelli$^{3,4}$\orcidlink{0000-0003-0148-3254},
Tommaso Ronconi$^{5}$\orcidlink{0000-0002-3515-6801},
\newauthor
Matteo Schulz$^{6,7}$\orcidlink{0009-0005-8184-0232},
Riccardo Murgia$^{8,9,10}$\orcidlink{0000-0002-2224-7704},
Jan Harms$^{6,7}$\orcidlink{0000-0002-7332-9806},
Tessa Baker$^{11}$\orcidlink{0000-0001-5470-7616},
Matteo Calabrese$^{12}$,
\newauthor
Carmelita Carbone$^{13}$\orcidlink{0000-0003-0125-3563},
Steven Cunnington$^{11}$\orcidlink{0000-0001-6594-107X},
Ian Harrison$^{14}$\orcidlink{0000-0002-4437-0770},
Konstantin Leyde$^{15}$\orcidlink{0000-0001-7661-2810},
\newauthor
Dounia Nanadoumgar-Lacroze$^{16}$\orcidlink{0009-0009-7255-8111}
\\
$^{1}$Marietta Blau Institute - Austrian Academy of Sciences, 1010 Vienna, Austria\\
$^{2}$INFN, Sezione di Roma, 1-00185 Roma, Italy\\
$^{3}$Observatoire de la Côte d’Azur, Laboratoire Lagrange, Bd de l’Observatoire, 06304 Nice, France\\
$^{4}$Department of Physics \& Astronomy, University of the Western Cape, Cape Town 7535, South Africa\\
$^{5}$INAF - Institute of Radioastronomy (IRA), Via Gobetti 101, 40129 Bologna, Italy\\
$^{6}$Gran Sasso Science Institute (GSSI), Viale F. Crispi 7, L'Aquila (AQ), I-67100, Italy\\
$^{7}$INFN - Laboratori Nazionali del Gran Sasso (LNGS), L'Aquila (AQ), I-67100, Italy\\
$^{8}$Dipartimento di Fisica, Universit\`a degli Studi di Cagliari, Cittadella Universitaria, 09042 Monserrato (CA), Italy\\
$^{9}$INFN, Sezione di Cagliari, Cittadella Universitaria, 09042 Monserrato (CA), Italy\\
$^{10}$INAF - Osservatorio Astronomico di Cagliari, Via della Scienza 5, 09047 Selargius (CA), Italy\\
$^{11}$Institute of Cosmology \& Gravitation, University of Portsmouth, Dennis Sciama Building, Portsmouth, PO1 3FX, UK\\
$^{12}$Astronomical Observatory of the Autonomous Region of Aosta Valley (OAVdA), Loc. Lignan, 39, I-11020 Nus, Italy\\
$^{13}$INAF - Institute of Space Astrophysics and Cosmic Physics (IASF), Via Corti 12, I-20133 Milano (MI), Italy\\
$^{14}$School of Physics and Astronomy, Cardiff University, CF24 3AA, UK\\
$^{15}$Center for Computational Astrophysics, Flatiron Institute, 162 5th Ave, New York, NY 10010\\
$^{16}$Institut de Física d’Altes Energies (IFAE), The Barcelona Institute of Science and Technology, Campus UAB, E-08193 Bellaterra, Barcelona, Spain
}

\pubyear{2026}

\begin{document}
\label{firstpage}
\pagerange{\pageref{firstpage}--\pageref{lastpage}}
\maketitle

\begin{abstract}
A new synergy between \acp{gw} and the study of the large-scale structure of the Universe is now emerging. Along this line of research, we combine simulated observations of stellar-origin black hole mergers and neutral hydrogen 21\,cm line intensity mapping to probe the expansion rate of the Universe through the distance-redshift relation. \Ac{gw} signals from binary black holes provide direct distance information, while neutral hydrogen intensity maps offer a tomographic view of the large-scale structure of the Universe. Using the 3-dimensional density fields of hydrogen as a redshift prior for \ac{gw} events, we explore a novel dark-sirens-like approach, here termed \textit{radio sirens}, to measure the late-time expansion history of the Universe. We study the performance of the next-generation \ac{gw} observatories, such as the Einstein Telescope, to ensure enough statistics and access to high-redshift data. On the other hand, future spectroscopic intensity mapping surveys with the SKA-Mid telescope are expected to trace the underlying dark matter distribution at large scales up to redshift $z\sim 3$. This combined methodology allows us to constrain the Hubble constant to $\sim 8$\% precision, using around 3,000 GW events with signal-to-noise ratios greater than 150. This corresponds to an improvement of around $90\%$ compared to not considering the information from the neutral hydrogen maps.
% Additionally, we show that if \ac{gw} sources trace the dark matter distribution, then assuming them to be homogenous and isotropic, as is currently done, will introduce a bias. \sm{final sentence about bias parameter}
\end{abstract}


%%%%%%%%%%%%%%%%%%%%%%%%%%%%%%%%%%%%%%%%%%%%%%%%%%

%%%%%%%%%%%%%%%%% BODY OF PAPER %%%%%%%%%%%%%%%%%%
\acresetall %restart acronyms "counts"
\section{Introduction}\label{sec:intro}

\Acp{gw} have recently joined the cosmological scene, acting as standard sirens. They provide a direct measurement of the luminosity distance to the source, circumventing the calibration problems of more traditional probes such as standard candles \citep{Schutz:1986gp, Holz:2005df}. Inference on cosmological parameters requires independent measurements of luminosity distance and redshift. Even though we have distance information, to extract cosmological information from \acp{gw}, we also need redshift information. 
The most straightforward way is represented by the so-called \textit{bright sirens} \citep{Markovic:1993cr, Dalal:2006qt, LIGOScientific:2017adf, Chen:2017rfc, Feeney:2018mkj, Palmese:2020, Mancarella:2024qle, Cozzumbo:2024vxw}, where the host galaxy is uniquely identified, and its redshift is measured electromagnetically. This requires that the gravitational event is accompanied by its \ac{em} counterpart. This latter needs at least one neutron star to be involved in the merger, and so far, we have only detected one such bright siren event \citep{LIGOScientific:2017adf}. 
Without an \ac{em} counterpart, \ac{gw} cosmology relies on inferring the redshift information from the source-frame mass distribution, the \textit{spectral siren} method \citep{Chernoff:1993th, Taylor:2012db, You:2020wju, Ye:2021klk,  Ezquiaga:2022zkx, Mastrogiovanni:2023emh, Ferraiuolo:2025evh, MaganaHernandez:2025cnu, Pierra:2026ffj, Tagliazucchi:2026dpr, Bertheas:2026odj}. In this case, a model of the astrophysical mass distribution is needed, whereas what is measured is the detector-frame mass of each event. 
In addition to the information from the mass modeling, constraints come from galaxy catalogs, where each galaxy represents a possible host of the \ac{gw} source.

In the context of \textit{dark sirens}, galaxy catalogs have been considered in two possible ways. The first is to use them to reconstruct a galaxy density profile, referred to as a \ac{los} redshift prior~\citep{Gray:2023wgj}, and then statistically allocate GW sources more likely in overdensities of this profile. This type of approach is usually referred to as the ``galaxy catalog'' method~\citep{Schutz:1986gp, Holz:2005df, MacLeod:2007jd,DelPozzo:2011vcw, Nishizawa:2016ood, Chen:2017rfc, LIGOScientific:2018gmd,DES:2019ccw,Gray:2019ksv,DES:2020nay,LIGOScientific:2019zcs,Finke:2021aom,LIGOScientific:2021aug,  Mancarella:2021ecn, Gair:2022zsa, Gray:2023wgj, Borghi:2023opd, Bom:2024afj, Borghi:2025nis}.
The second is to use summary statistics for the \ac{gw}s and galaxies, such as the cross and auto angular power spectra \citep{Oguri:2016dgk, Bera:2020, Mukherjee:2020hyn, Diaz:2021pem, Scelfo:2021fqe, Mukherjee:2022afz, Mali:2024wpq, Ferri:2024amc, Pedrotti:2025tfg, SantiagodeMatos:2025iyj, Dalang:2024gfk}. This method is typically referred to as the ``cross-correlation'' method.
These two methods are actually two sides of the same method, and can be unified under some assumptions \citep{Cheng:2026atn}.

In this paper, we investigate the synergy between \ac{bbh} standard sirens as would be measurable with the \ac{et} and a \ac{hi} \ac{im} survey with the \ac{skao}~\citep{SKA}. We use a hybrid approach between the ``cross-correlation'' and ``galaxy catalog'' methods, where the density contrasts of \ac{hi} calculated from the \ac{lss} are used to build the \ac{los} redshift priors for the \ac{gw} signals.
Via the measurement of the redshifted 21\,cm line, SKA-Mid, the SKAO radio telescope in South Africa,  can potentially map, in single-dish mode, the \ac{lss}, covering an unprecedented sky area of 20,000\,deg$^2$ up to $z\sim 3$~\citep{SKA:2018ckk}. 
\ac{hi} \ac{im}, with the one-to-one correspondence between the redshift of emission of the 21\,cm signal and the observed frequency, offers a spectroscopic view of the matter distribution over the redshift interval in which \ac{gw} distance measurements are available, providing an advantage over photometric galaxy catalog \textit{dark siren} methods. This emerging technique is advancing fast thanks to purpose built telescopes as CHIME\footnote{\url{https://chime-experiment.ca/}} and HIRAX\footnote{\url{https://hirax.ukzn.ac.za/}} and to a large observational campaign carried out with the SKA-Mid precursor MeerKAT\footnote{\url{https://www.sarao.ac.za/science/meerkat/}}. MeerKAT's Large Area Synoptic Survey (MeerKLASS) \citep{MeerKLASS:2017vgf} will reach a survey area of ${\sim}10{,}000{\rm deg}^2$, up to $z\,{\sim}\,1.4$ \citep{Cunnington:2025sdr}, and the field has already achieved successful calibration \citep{Wang:2020lkn} and cosmological detections \citep{2023MNRAS.518.6262C,MK25}.

Previous works~\citep{Scelfo:2021fqe, Zazzera:2025ord} focused on \ac{gw} and \ac{hi} summary statistics, such as the angular power spectrum ~\citep[see also][\textit{in preparation}]{Schulz:InPrep}. Here we focus on connecting \ac{lss} to the number density of \ac{cbc} mergers: using the 3-dimensional density fields of \ac{hi} measured with radio telescopes to model the \ac{los} redshift prior for \ac{gw} events, we explore a dark-sirens-like approach, that we term \textit{radio sirens}, to measure the late-time expansion history of the Universe. 
%\footnote{We chose to use \textit{radio sirens} to pinpoint that the redshift information comes from measurements of the 21\,cm \ac{hi} line in the radio band.}
A similar approach combining \acp{gw} and \ac{hi} temperature maps is also being explored in~\cite[][\textit{in preparation}]{Nanadoumgar-Lacroze:InPrep}. 

Our goal is to isolate and quantify the constraining power of \ac{hi} data. To this aim, we do not include any information from the \ac{bbh} mass spectrum. This allows us to avoid any dependence on a specific mass model. Our primary result is therefore a comparison of the $H_0$ posterior obtained with and without incorporating \ac{hi} maps in the \ac{los} modeling at the inference stage. 

The paper is organized as follows. In Section~\ref{sec:methods}, we describe the construction of the mock \ac{hi} maps and the \ac{gw} simulations, and the hierarchical Bayesian analysis framework. In Section~\ref{sec:results}, we
present the inferred posteriors for the Hubble parameter under different assumptions and setups. In Section~\ref{sec:disc_and_concl}, we discuss the implications, limitations, and future extensions of the proposed method.

\section{Methods}\label{sec:methods}

It can be assumed that both the \ac{hi} field~\citep{Santos:2015gra, Bull:2015stt} and \ac{bbh} mergers~\citep{Scelfo:2018sny, Pedrotti:2025tfg} are biased tracers of the \ac{dm} density field. This means that we can generally express the density contrast of a given tracer of the underlying \ac{dm} density field as:
\begin{equation}\label{eq:delta_tracer}
\delta_{\rm{tracer}}(z, {\rm{RA}}, {\rm{Dec}})=b_{\rm{tracer}}(z)\delta_{\rm{DM}}(z, {\rm{RA}}, {\rm{Dec}}),    
\end{equation}
where $\delta_{\rm tracer}$ is the density contrast of the tracer, being it either \ac{hi} or \acp{gw}, and $\delta_{\rm DM}$ represents the \ac{dm} density contrast. The couple RA and Dec define the direction in the sky. $b_{\rm tracer}$ is the bias parameter, with $b_{\textrm{H\textsc{i}}}$ for \ac{hi} and $b_{\rm GW}$ for \acp{gw}. Note that we are working under the simplified assumption that the bias is a function of redshift only. 

If \ac{bbh} mergers preferentially occur in overdense regions, then, since we do not have direct access to \ac{dm} field distribution, the observed \ac{hi} density can be used as an informative prior on the redshift of \ac{gw} events. This gives us the distance-redshift relation needed for cosmological inference. Building on this idea, we develop a proof-of-concept \textit{radio siren} framework and test how much \ac{hi} tomography can improve constraints on cosmological parameters, particularly the Hubble constant. This is done in the context of the planned \ac{3g} \ac{gw} detectors, with the \ac{et} exemplifying the future gravitational interferometers' capabilities, to ensure sufficient statistics and redshift reach. In this section, we begin by explaining how we build simulated radio observations of \ac{hi} intensity maps, then how we connect them to the \ac{cbc} merger rate, and how we simulate the mock \ac{gw} data. Finally, we describe how the simulated radio and \ac{gw} data are combined within a coherent analysis framework to infer cosmological parameters.

\subsection{Building DM maps}\label{sec:DM_maps}

Our starting point is the simulation of the \ac{dm} distribution. Given that SKA-Mid frequency coverage will allow tomography of the $21$\,cm line of \ac{hi} up to $z\sim 3$ and \ac{et} can trace \acp{bbh} up to even $z\sim 100$~\citep{Branchesi:2023mws, ET:2025xjr}, we set our analysis for $z\in [0, \sim 3]$.

We construct mock \ac{lss} maps starting from \ac{dm} halo catalogs derived from the \ac{DEMNUni} set of simulations~\citep{Castorina:2015bma, Carbone:2016nzj, Parimbelli:2022} in their high-resolution configuration \citep[HR-\ac{DEMNUni},][]{Hernandez-Molinero:2023jes, Verza:2024}. These simulations have been produced using the Gadget-3 code~\citep{Springel:2005mi}. They are characterized by a Planck 2013~\citep{Planck:2013pxb} baseline flat $\Lambda$\,\ac{cdm} cosmology with different values of the total neutrino mass and of the parameters characterizing the \ac{de} \ac{eos}. Our model considers only the massless neutrino flat \ac{cdm} case and a box with side length of $500$\,Mpc\,$h^{-1}$ in comoving coordinates, with $2048^3$ \ac{dm} particles. A full-sky lightcone with three-dimensional geometry from redshift $z=0$ to redshift $z\approx8$ is then built, corresponding to a total comoving volume of approximately $3\times10^3$\,Gpc$^{3}$\,$h^{-3}$.
The approach used to build this lightcone, originally developed for weak lensing applications~\citep{Carbone2008,Calabrese2015,Hilbert2020}, preserves the continuity of the gravitational potential across transverse directions.
The lightcone volume is partitioned into concentric spherical shells of fixed comoving thickness, within which simulation outputs are coherently translated and rotated. Around $39$ billion haloes are identified through the \ac{fof} algorithm.

From the \ac{dm} halo catalogs, we construct sky maps in redshift shells by counting the number of \ac{dm} halos per pixel and converting these counts into overdensity maps:
\begin{equation}\label{eq:map_construction}
{\rm map}(z,{\rm pix}_i) =
\frac{N(z,{\rm pix}_i) - \bar{N}_{\rm pix}(z)}{\bar{N}_{\rm pix}(z)},
\end{equation}
with
\begin{equation}\label{eq:average_dm_halos_per_pix}
\bar{N}_{\rm pix}(z) =
\frac{1}{n_{\rm pix}}
\sum_i N(z,{\rm pix}_i)
\end{equation}
where $N(z,{\rm pix})$ is the number count at a given redshift shell and pixel, $\bar{N}_{\rm pix}(z)$ is the averaged number count per redshift shell, and $n_{\rm pix}$ is the total number of pixels, given the adopted resolution (see Sec.~\ref{sec:gw_sim}). pix$_i$ represents the direction in the sky (RA, Dec). We adopt a Healpy~\citep{Gorski:2004by, Zonca:2019vzt} pixelization with \texttt{nside}=16, which roughly corresponds to a resolution of $\sim 13$\,deg$^2$ per pixel. The binning in redshift, instead, follows approximately an equally-spaced distribution in the logarithm of the scale factor $a=(1+z)^{-1}$. This stems from the spherical shells partition of the lightcone as described above.
%It has been constructed following the redshift spacing of the snapshots used to construct the light cone in the HR-DEMNUni simulation.

In building the \ac{dm} density maps, we make the simplified assumption that haloes describe the \ac{dm} background field. Even though this assumption is strong, we have verified that at the scales we inspect and for the exploratory purpose of this work, it does not affect our final conclusions.

\subsection{Radio observations of neutral hydrogen}\label{sec:hi_maps}

In this section, we build mock \ac{hi} radio observations starting from the DM maps generated in Sec.~\ref{sec:DM_maps}.

The 21\,cm line of \ac{hi} rest-frame frequency $\nu_{21}$ is redshifted to lower frequencies due to cosmic expansion according to 
\begin{equation}\label{eq:hi_rest_freq}
    \nu_{\mathrm{obs}} = \nu_{21}/(1+z) \,,~~ {\rm with }~~\nu_{21}=1420~{\rm MHz}
\end{equation}
so that each instrumental frequency band can be directly mapped to a corresponding redshift interval.

The 21 cm line emission from \ac{hi} is a faint signal and it can be observed in emission from individual sources only at low redshift. By giving up on resolution and integrating the signal from multiple objects, \ac{im} allows instead to map the \ac{lss} at unprecedented depths (e.g. ~\citealt{Battye_2013,Kovetz:2017agg}).
\ac{im} surveys with the radio telescopes are thus capable of tracing the underlying \ac{dm} distribution at large scales up to high redshift.

\begin{figure}
    \centering
    \includegraphics[width=0.9\columnwidth]{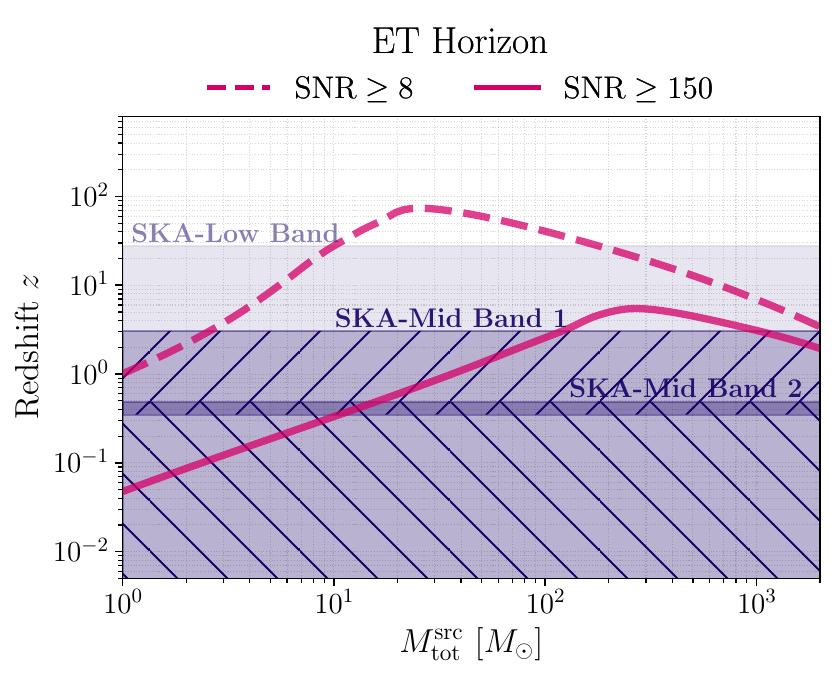}
    \caption{Horizon plots (\textit{in magenta}) for the \ac{et} observations compared to the \ac{skao} reach in redshift. The horizon represents the maximum redshift for \ac{cbc} equal-mass and optimally oriented \ac{gw} sources observed with a threshold \ac{snr}. Here we compare the standard SNR = 8 threshold (\textit{dashed line}), with the one we use throughout this work, i.e. SNR = 150 (\textit{solid line}). On the \textit{x-axis} is the total source-frame mass of the binary, while on \textit{y-axis} is the redshift of the merger. We used the triangular configuration for \ac{et} with the full cryogenic sensitivity curve available \href{https://apps.et-gw.eu/tds/?r=18213}{here}. The \ac{skao} redshift ranges are highlighted horizontally in \textit{purple}: there are the two bands of SKA-Mid, band 2 ($950-1760$\,MHz) and band 1 ($350-1050$\,MHz), and the high-redshift SKA-Low band ($50-350$\,MHz). The intersection region of the two SKA-Mid bands is marked with two different hatches. In this work, we focus on both bands of SKA-Mid.}
    \label{fig:horizon}
\end{figure}

The \ac{skao} will consist of two instruments, SKA-Low in Australia covering the band $50-350$\,MHz and SKA-Mid in South Africa, observing in band 2 ($950-1760$\,MHz) and band 1 ($350-1050$\,MHz). The \ac{skao} observational bands are illustrated in Fig.~\ref{fig:horizon}, along with their corresponding redshift ranges. The figure highlights the synergy between \ac{skao} and \ac{et} investigated in this work. Further details about \acp{gw} are provided in Sec.~\ref{sec:gw_sim}.

Using SKA-Mid in single-dish mode, it will thus be possible to measure the \ac{lss} up to
$z\sim 3$~\citep{SKA:2018ckk}, covering a much larger cosmological volume than possible with current SKA precursor observations.

As a first approximation, we assume a one-to-one correspondence between the \ac{hi} distribution and the \ac{dm} field. In fact, as shown in Eq.~\ref{eq:delta_tracer}, we can set the \ac{hi} bias parameter $b_{\textrm{H\textsc{i}}}$ to one with respect to the \ac{dm} density contrast $\delta_{\rm DM}$. Moreover, we make the simplified assumption that we can work with full-sky coverage.

We obtain the \ac{hi} intensity maps from Eq.~\ref{eq:map_construction}, setting $b_{\textrm{H\textsc{i}}}=1$, using the same sky resolution and redshift binning of the \ac{dm} maps. This translates to a minimum frequency resolution above the MHz, larger than the instrumental capabilities of SKA-Mid. We show example maps at two different reference redshifts later on in Fig.~\ref{fig:sky_maps_plus_gws} (Sec.~\ref{sec:gw_sim}) in connection with \ac{gw} observations.

Note that what is effectively observed with \ac{hi} \ac{im} surveys is the temperature brightness in each pixel, which depends on the total \ac{hi} density as a function of redshift, $\Omega_{\textrm{H\textsc{i}}}$~\citep{Battye_2013, Crighton2015, P_roux_2020}. We assume this function is known and directly use the \ac{hi} density fluctuations for the analyses presented here.

Another approximation we make is not to explicitly include the chromatic beam response that would result from using SKA-Mid in single-dish mode~\citep{Santos:2015gra, SKA:2018ckk}. Given that we work with \texttt{nside}$=16$ (see Sec.~\ref{sec:hi_maps}), the worst smoothing due to the beam (associated with the lower SKA-Mid frequency available and corresponding to redshift 3 in terms of the 21\,cm line) is lower than the pixel resolution and therefore is not relevant for our settings.

Finally, we do not account for any systematics in the \ac{hi} maps or for the effect of foreground cleaning~\citep[e.g.][]{Spinelli:2021emp}. Some of the \ac{hi} systematics and limited coverage of the maps will be addressed in~\citep[][\textit{in preparation}]{Nanadoumgar-Lacroze:InPrep}.

These assumptions define an optimistic baseline scenario. In realistic applications, deviations from these conditions are expected and may degrade or bias cosmological constraints. Quantifying all these effects in more detail is deferred to future work.

\subsection{Mock \ac{gw} data}\label{sec:gw_sim}

To connect \acp{gw} with the underlying \ac{dm} field, we need to parametrize the \ac{cbc} merger rate as a function of the \ac{dm} density contrast.
We model the merger rate as the number of \ac{cbc} sources, $N_{\rm CBC}$, per redshift $z$, detector-frame time $t$ and sky position $\Omega$ as
\begin{equation}
\frac{{\rm d}N_{\rm CBC}}{{\rm d}z\,{\rm d}t {\rm d}\Omega}
=
\frac{{\rm d}N_{\rm CBC}}{{\rm d}M_{\textrm{BH}}\,{\rm d}t_{\rm src}}
\frac{1}{1+z}
\frac{{\rm d}M_{\textrm{BH}}}{{\rm d}z\,{\rm d}\Omega},
\label{eq:rate}
\end{equation}
where $M_{\textrm{BH}}$ is the mass of available compact objects in binaries, and $t_{\rm src}$ is the time in the source frame. The first term describes the source-frame merger rate per available \ac{bh} mass, the second term is the distribution of binaries in redshift and angular position.

We parameterize the source-frame merger-rate term using a modified Madau-Dickinson model~\citep{Madau:2014bja},
\begin{equation}\label{eq:madau}
\begin{aligned}
\frac{{\rm d}N_{\rm CBC}}{{\rm d}M_{\rm BH}\,{\rm d}t_{\rm src}}
&=
R_{\rm BH}\,\psi(z\mid \gamma,\kappa,z_p)\\
&=R_{\rm BH}\,\ \left[1+ (1+z_p)^{-\gamma-\kappa} \right] \frac{(1+z)^{\gamma}}{1+ \left[ \frac{1+z}{1+z_p} \right]^{\gamma +\kappa}}
\end{aligned}
\end{equation}
where $\psi(z\mid \gamma,\kappa,z_p)$ is the Madau--Dickinson redshift evolution, and $R_{\rm BH}$ is the rate of mergers per year per solar mass available of compact objects. The overall merger rate acts as a multiplicative factor for the number density of \ac{bbh} mergers. $\psi(z\mid \gamma,\kappa,z_p)$ depends on the slope before, $\gamma$, and after, $\kappa$, of the $z_p$ parameter, which quantifies the turning point between the two power laws and identifies the peak of the \ac{cbc} merger rate. The choice of parameters in this work is specified in Tab.~\ref{tab:priors_table} (see also App.~\ref{app:gw_simulation}).

The distribution of \ac{bh} mass is factorized in terms of comoving mass density, namely
\begin{equation}
\frac{{\rm d}M_{\textrm{BH}}}{{\rm d}z\,{\rm d}\Omega}
=
\frac{{\rm d}M_{\textrm{BH}}}{{\rm d}V_c}
\frac{{\rm d}V_c}{{\rm d}z\,{\rm d}\Omega}
=
\rho_{\rm BH}(z,\Omega) \frac{{\rm d}V_c}{{\rm d}z\,{\rm d}\Omega},
\label{eq:rate0}
\end{equation}
and using linear order perturbations, we can write 
\begin{equation}
    \rho_{\rm BH}(z,\Omega) \approx \bar{\rho}_{\rm BH} [1+b_{\rm GW}(1+z)^{\alpha_{\rm GW}} \delta_{\rm DM}(z,\Omega)].
    \label{eq:bias}
\end{equation}
In the above equation, we have written the comoving density of available \acp{bh} in terms of an average density $\bar{\rho}_{\rm BH}$, the \ac{dm} density contrast $\delta_{\rm DM}(z,\Omega)$  and multiplicative bias parameter parametrized as in \cite{Mukherjee:2020hyn, Diaz:2021pem, Dalang:2023ehp}. Eq.~\ref{eq:bias} encapsulates the key physical assumption of the model: that the spatial distribution of the mergers is modulated by the underlying matter density field. The bias parameters $(b_{\rm GW}, \alpha_{\rm GW})$ encode how strongly and in what way the \ac{cbc} merger rate traces the \ac{lss} (see for example~\citealt{Pedrotti:2025tfg}).

In this exploratory work, we assume that \ac{gw}s perfectly trace the \ac{dm} distribution and therefore  $b_{\rm GW}=1, \alpha_{\rm GW}=0$. In a more general scenario, we can still parametrize $b_{\textrm{H\textsc{i}}}(z)$ and marginalize over its parameters' uncertainties.

Before proceeding, let us argue on how this method relates to current methodologies for dark sirens cosmology. If we combine Eq.~\ref{eq:madau} with Eq.~\ref{eq:rate0} and Eq.~\ref{eq:bias}, we can note that the usual binary black merger rate per comoving volume can be defined as $R_0=R_{\rm BH} \bar{\rho}_{\rm BH}$. In this sense, $R_0$ is the merger rate for a homogenous universe. By further assuming that the universe is perfectly homogenous and isotropic ($\delta_{\rm DM}(z,\Omega)=\delta_{\rm GW}(z,\Omega)=0$), then we can see that the rate in Eq.~\ref{eq:rate} collapses to the classical rate model used for dark sirens analyses \citep{Mastrogiovanni:2021wsd, Mastrogiovanni:2022ykr, Mastrogiovanni:2023zbw}.

In this work, we consider \Ac{3g} \ac{gw} detectors as they will probe compact binary mergers across a much larger redshift range than current instruments~\citep{Branchesi:2023mws, ET:2025xjr}. In particular, the \ac{et} is expected to detect \acp{cbc} out to very high redshift, opening the possibility of tracing the cosmic expansion well beyond the local Universe. This makes \ac{3g} \ac{gw} observatories particularly promising for standard siren cosmology. Specifically, we investigate the synergy between \ac{et} for the \ac{bbh} detection on the gravitational side and \ac{skao}.

In Fig.~\ref{fig:horizon}, we show the horizon for the \ac{et} observations compared to the \ac{skao} reach in redshift. The horizon represents the maximum redshift for equal-mass, optimally oriented \ac{cbc} \ac{gw} sources observed with a threshold \ac{snr}. Here we compare the standard \ac{snr} = 8 threshold with the one we use throughout this work, i.e., \ac{snr} = 150. We used the triangular configuration for ET with the full cryogenic sensitivity curve as presented in~\cite{Branchesi:2023mws}\footnote{The cryogenic curve for the 10\,km \ac{et} is publicly available \href{https://apps.et-gw.eu/tds/?r=18213}{here}.}. The same configuration has been used in all the analyses presented in this paper. 

\begin{figure*}
    \centering
    
    % ---------- First Row (2 plots) ----------
    \begin{subfigure}{0.48\textwidth}
        \centering
        \includegraphics[width=\linewidth]{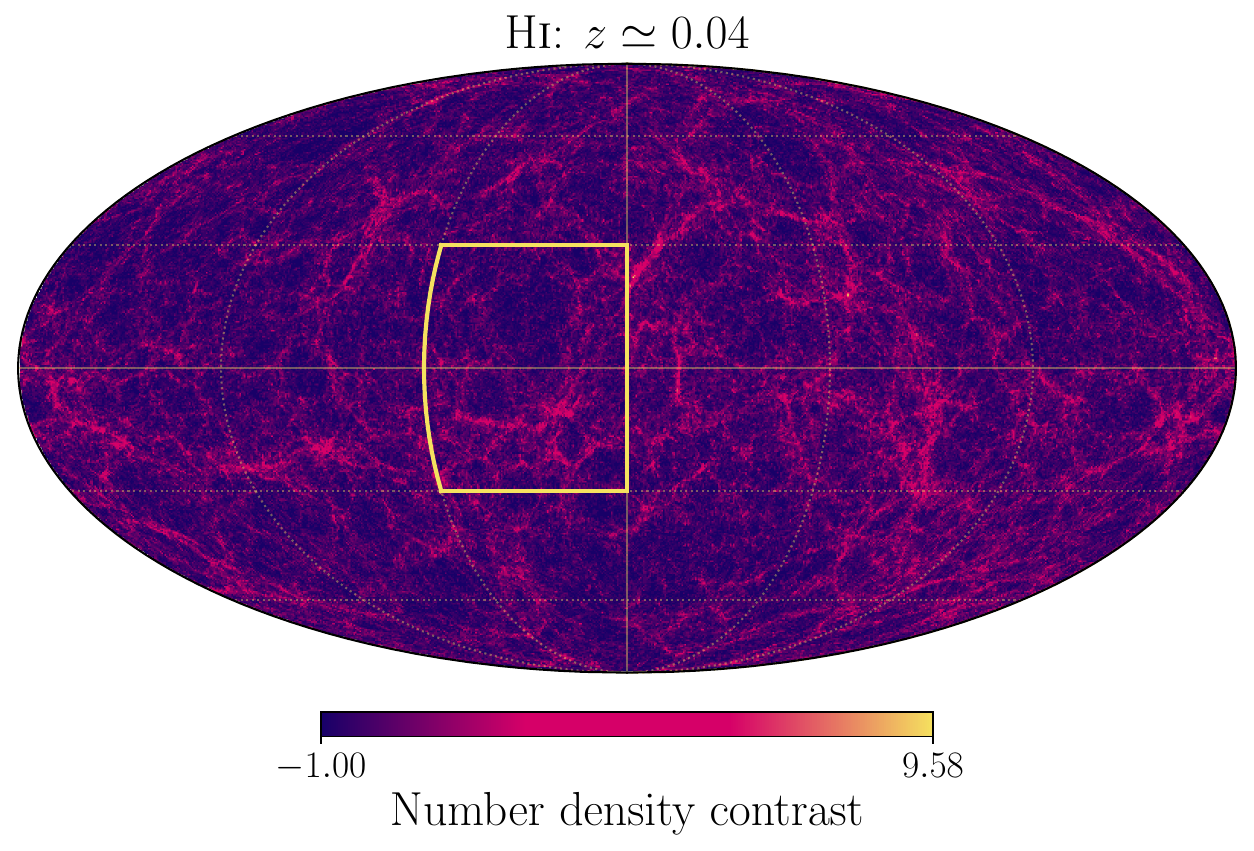}
    \end{subfigure}
    \hfill
    \begin{subfigure}{0.48\textwidth}
        \centering
        \includegraphics[width=\linewidth]{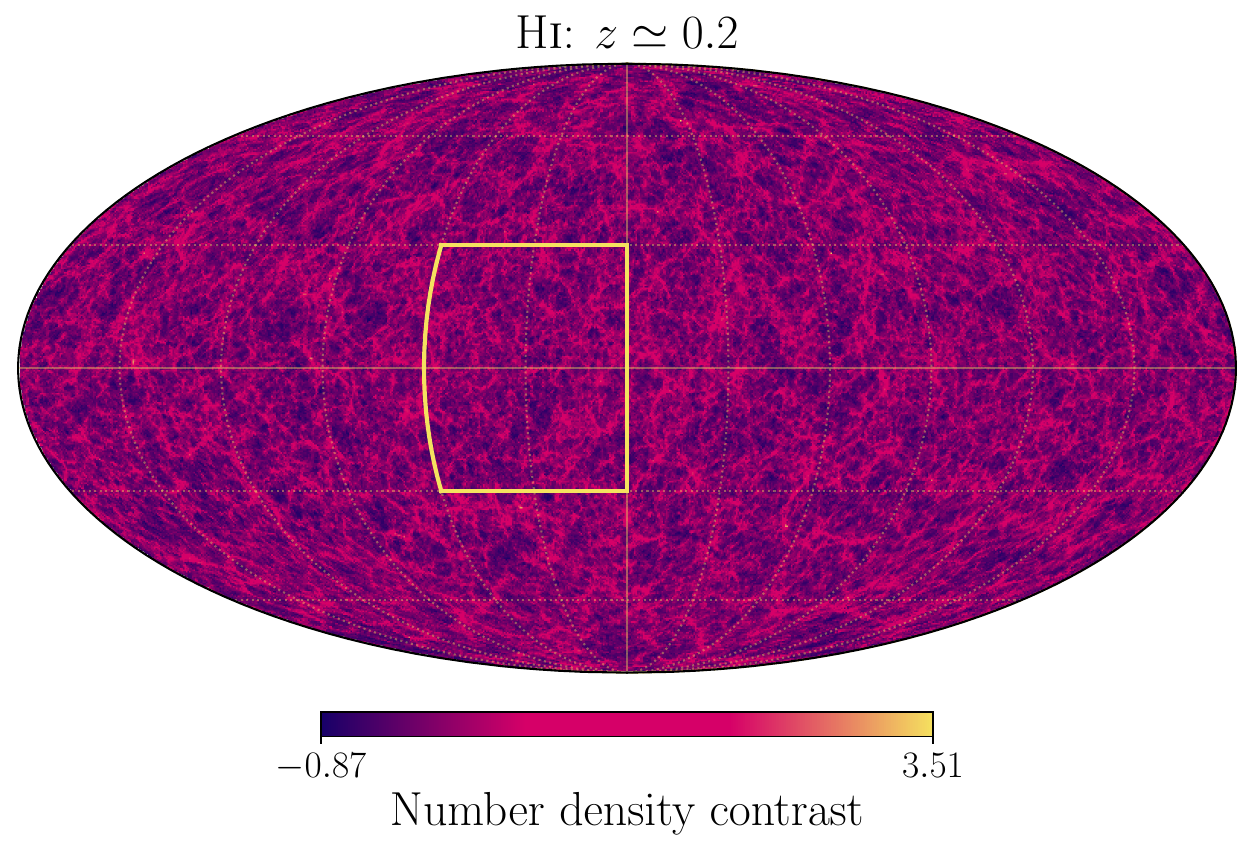}
    \end{subfigure}
    
    \vspace{0.3cm}
    
    % ---------- Second Row (2 plots) ----------
    \begin{subfigure}{0.48\textwidth}
        \centering
        \includegraphics[width=\linewidth]{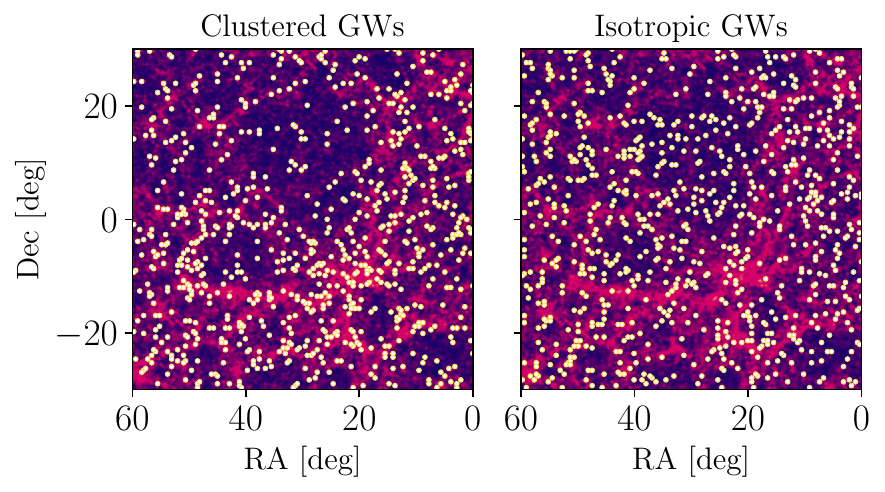}
    \end{subfigure}
    \hfill
    \begin{subfigure}{0.48\textwidth}
        \centering
        \includegraphics[width=\linewidth]{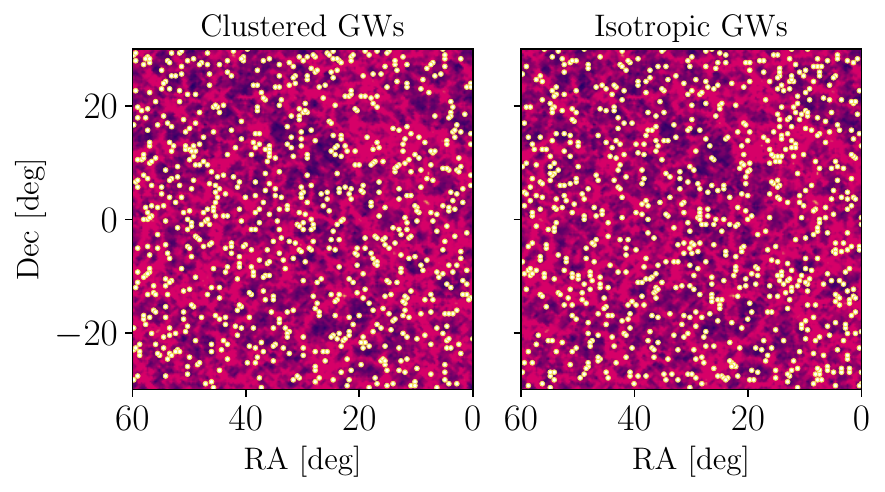}
    \end{subfigure}
    
    \caption{Sky maps of the \hi\ density contrast and simulated \ac{gw} event distributions at two representative redshifts. \textit{Top panels}: Mollweide projections of the \hi\ density contrast at $z\simeq 0.04$ (left) and $z\simeq 0.2$ (right). The yellow boxes indicate the sky region shown in the zoomed-in panels below. Color bars denote the amplitude of the density contrast. \textit{Bottom panels}: Zoom-in of the boxed regions, comparing clustered \ac{gw} events (left subpanels), which trace the underlying \hi\ distribution, with isotropically distributed \ac{gw} events (right subpanels). We increased the number of \ac{gw} sources to 10k per map and the resolution of the map (using \texttt{nside}=128) for visualization purposes. While at low $z$ the difference between clustered and isotropically distributed \ac{gw} events is more pronounced, this distinction progressively diminishes at higher redshifts, where the two distributions become increasingly similar, according to the variation of the \ac{dm} density contrast as shown in Fig.~\ref{fig:los_profile}.}
    \label{fig:sky_maps_plus_gws}
\end{figure*}

We simulate $10^4$ \ac{bbh} events. For our choice of rate parameters for Eq.~\ref{eq:madau}, this corresponds to one year of observations. Note that this is one order of magnitude less than the typical amount of observations per year expected for \ac{et}~\citep{ET:2019dnz, Branchesi:2023mws} as a consequence of rate parameters (see Tab.~\ref{tab:priors_table}). Our choice was motivated by the need to concentrate the bulk of the events in the redshift range covered by the \ac{hi} maps, i.e. $z\in [0, \sim3]$. In the following, we provide a brief description of how the parameters characterizing the \ac{gw} events have been sampled. We refer to App.~\ref{app:gw_simulation} for details.

We sample the masses uniformly in the detector frame\footnote{This is done for consistency in the hierarchical Bayesian analysis framework, see Sec.~\ref{sec:hbi} for more details. We note that it implies a non-trivial mass distribution in the source frame. However, as no explicit mass-spectrum models are incorporated in our analysis, adopting a distribution that is uniform in detector-frame mass provides a convenient and practical assumption.} in the mass range $[5M_{\odot}, 300M_{\odot}]$. Sky position and distance are chosen after the \ac{dm} maps. All the other parameters are sampled uniformly in their range (see App.~\ref{app:gw_simulation} and Tab.~\ref{tab:gw_ranges} therein for details). We then retain the subset of events that is detected by \ac{et} in its triangular configuration with the full cryogenic sensitivity curve as in~\cite{Branchesi:2023mws} with an \ac{snr} threshold of 150. % This yields approximately $3100$ detected events in both the clustered and isotropic scenarios. 
The high \ac{snr} threshold is needed to ensure that our events are well-behaved under the Fisher analysis, which is used for their \ac{pe}. For each event, in fact, \ac{pe} is performed in the Fisher-matrix approximation using \texttt{GWFish}~\citep{Dupletsa:2022scg, Dupletsa:2024gfl}. To ensure the samples lie within their physical range, we post-processed the Fisher results with priors using sampling from the truncated multivariate Gaussian as described in~\cite{Dupletsa:2024gfl}. The resulting catalogs span a broad range of luminosity distances, and the 90\% sky-localization areas generally increase with distance. Around $15\%$ of events are localized on angular scales less than or comparable to the chosen map pixel area. Given the adopted Healpy pixelization (\texttt{nside}=16, see Sec.\ref{sec:hi_maps}), we broadly match the sky-localization precision of the simulated \ac{gw} events (see App.~\ref{app:gw_simulation} and Fig.~\ref{fig:sky_loc_distance_scatter} therein).

The maps of \ac{dm} density contrast (see Section~\ref{sec:DM_maps} for details) are used to populate the universe with \ac{gw} sources using Eqs.~\ref{eq:rate0} and~\ref{eq:bias}. Since we want to probe the effect of clustering for \ac{gw}s, we construct two catalogs of \ac{gw} detections:
\begin{enumerate}
    \item \textbf{Clustered \acp{gw}:} \ac{bbh} events are distributed according to the same sky probability field traced by the \ac{dm} maps.
    \item \textbf{Isotropic \acp{gw}:} \ac{bbh} events are distributed uniformly in the sky, instead of explicitly following the \ac{dm} clustering. %giving up on the information coming from the correspondence between the source population and the \ac{hi} clustering prior. 
    %In other words, the \ac{gw} bias parameter is set to $0$ and \acp{gw} do not trace the \ac{dm} distribution.
\end{enumerate}

Under our choice of \ac{snr} threshold, this yields approximately $3100$ detected events in both the clustered and isotropic scenarios. In Fig.~\ref{fig:sky_maps_plus_gws}, we show sample sky maps of the \ac{hi} density contrast and simulated \ac{gw} event distributions at two representative redshifts, low and relatively high: $z\simeq 0.04$ (left panels) and $z\simeq 0.2$ (right panels), respectively. The yellow boxes indicate the sky region shown in the zoomed-in panels below. In the zoomed insets, we compare clustered \ac{gw} events (left subpanels), which trace the underlying \ac{dm} distribution, with isotropically distributed \ac{gw} events (right subpanels). 

While at low $z$ the difference between clustered and isotropically distributed \ac{gw} events is more pronounced, this distinction progressively diminishes at higher redshifts, where the two distributions become increasingly similar, according to the variation of the \ac{dm} density contrast. These directional redshift-dependent fluctuations are the key ingredient to make the \ac{hi} prior informative. For visualization purposes, a higher-resolution pixelization is used in the sky maps of Fig.~\ref{fig:sky_maps_plus_gws}. Additionally, we also increased the number of \ac{gw} sources to 10k for each shown redshift slice. 

\subsection{The \ac{hbi} framework}\label{sec:hbi}
The aim of this work is to constrain cosmological parameters using \acp{gw} as \textit{radio sirens}: we use the luminosity distance information coming from gravitational mock measurements and the redshift information from the simulations of \ac{hi} maps. In order to do so, we work in the \ac{hbi} framework. This is a two-level inference analysis combining posterior samples coming from individual \ac{gw} events with population-level information. 

Given $N_{\rm obs}$ \ac{gw} events, collectively specified as $\{x\}$, each of which is described by its own set of parameters $\theta$, the hierarchical likelihood that describes the probability of observing the specific dataset in a given time $T_{\rm obs}$, and given the set of hyperparameters $\Lambda$, is~\citep{Mandel:2018mve, Vitale2020aaz, Mastrogiovanni:2021wsd, Mastrogiovanni:2023emh, Mastrogiovanni:2023zbw}:
\begin{equation}
    \label{eq:hierarchical_lkh}
    \begin{aligned}
    \mathcal{L}(\{x\}|\Lambda) \propto \exp{\left[-N_{\rm exp}(\Lambda) \right]}\prod_{i=1}^{N_{\rm obs}} T_{\rm obs} \int \mathcal{L}_{\rm GW}\left(x_i|\theta, \Lambda \right)\frac{\dd N_{\rm CBC}}{\dd t \dd \theta}(\Lambda)\dd\theta,
    \end{aligned}
    \end{equation}
where, we have the single-event likelihood $\mathcal{L}_{\rm GW}\left(x_i|\theta, \Lambda \right)$ weighted by the population-level probability $\frac{\dd N_{\rm CBC}}{\dd t \dd \theta}(\Lambda)$ term as explained in Eq.~\ref{eq:rate}-\ref{eq:bias}. The hyperparameters $\Lambda$ govern the population-level model, describing the distribution of \ac{gw} events as a function of redshift and sky position. In this study, we identify 5 population-level parameters that are detailed with their injection values and priors in Tab.~\ref{tab:priors_table}.

\begin{table}
	\centering
	\caption{Uniform prior distributions adopted for the cosmological and rate parameters in the \texttt{icarogw} analysis. $\mathcal{U}[a,b]$ denotes a uniform distribution between $a$ and $b$. The last column reports the injected values used in our simulations.}
	\label{tab:priors_table}
    \renewcommand{\arraystretch}{1.1}
	\begin{tabularx}{\columnwidth}{l X X }%{\centering\arraybackslash} X}
    \toprule
        \textbf{Parameter} &\textbf{Prior range} &\textbf{Injected value}\\
    \midrule
        \rowcolor{SKABlue!10}
		$H_0\,[\mathrm{km}\,\mathrm{s}^{-1}\,\mathrm{Mpc}^{-1}]$ &${\mathcal{U}}[10, 240]$ &$67.7$\\
		$\Omega_{m,0}$ & ${\mathcal{U}}[0.1, 0.9]$ &$0.308$ \\
        \rowcolor{SKABlue!10}
		$\gamma$ & ${\mathcal{U}}[0, 12]$  &$2.7$\\
		$\kappa$ & ${\mathcal{U}}[0, 12]$  &$6$\\
        \rowcolor{SKABlue!10}
        $z_p$ & ${\mathcal{U}}[0, 4]$  &$1$\\
		\bottomrule
	\end{tabularx}
\end{table}

Additionally, we have the term dependent on the expected number of events $N_{\rm exp}$ in the observing time interval $T_{\rm obs}$. This term is necessary to properly account for \ac{gw} selection effects. More details about this framework, and \textsc{icarogw}, the software tool used for this study, are given in \cite{Mastrogiovanni:2023emh}. Some details on the numerical implementation of the \ac{hbi} framework in \textsc{icarogw} are also given in App.~\ref{app:icaro}.

\begin{figure}
    \centering    \includegraphics[width=0.9\columnwidth]{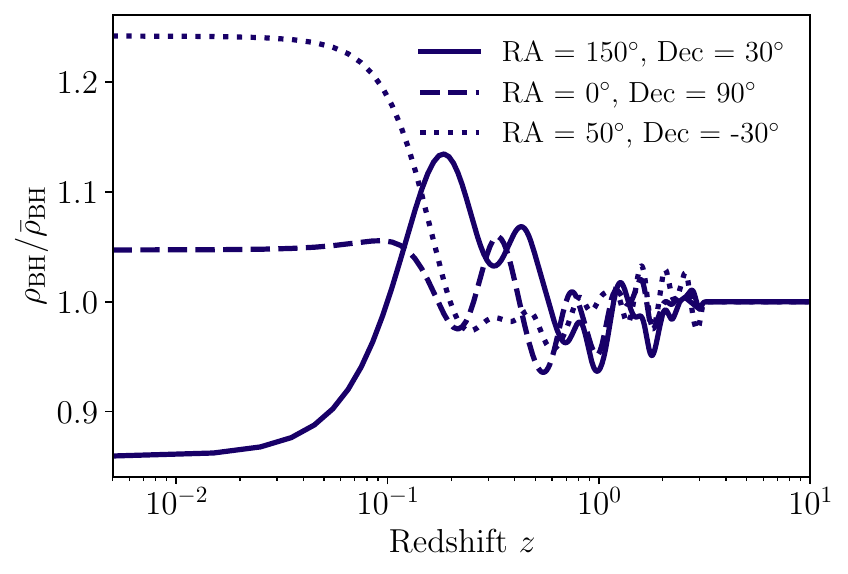}
    \caption{Distribution of $\rho_{\rm BH} / \bar{\rho}_{\rm BH}$
    along three different lines of sight identified by their right ascension RA and declination Dec (specified in the legend). The variation is more pronounced at low redshift and becomes increasingly uniform toward higher redshift, approaching homogeneity by $z\sim 3$, the maximum redshift covered by the \hi\ maps. The apparent flattening at low redshift is a binning artifact: the input redshift grid is sparse at low $z$.} %\UD{Comment on the flat distribution at low $z$ related to bin size.}\TR{do we have a LoS with higher overdensity at low redshift? something above 1.2 to balance the plot.}\sm{replace with density of BHs on y-axis}}
    \label{fig:los_profile}
\end{figure}

The key assumption of our analysis is that in the modeling of the \ac{los} redshift prior, we use the simulated \ac{hi} maps as the observable tracer of the \ac{dm} field appearing in Eq.~\ref{eq:bias}. We show the quantity $\rho_{\rm BH}/\bar{\rho}_{\rm BH}$ in Fig.~\ref{fig:los_profile} as a function of redshift for different directions. The variation is more pronounced at low redshift and becomes increasingly uniform toward higher redshift, approaching homogeneity by $z\sim 3$, the maximum redshift covered by the \ac{hi} maps. The apparent flattening at low redshifts is a binning artifact.

Conceptually, the \textit{radio-siren} framework can be interpreted as a low spatial resolution generalization of galaxy-catalog dark sirens. Instead of assigning discrete host probabilities to resolved galaxies, we use the large-scale density field traced by \ac{hi} to construct a redshift prior along each line of sight. Using \ac{hi} maps instead of discrete galaxy catalogs circumvents issues associated with heterogeneous, flux-limited surveys and mitigates the completeness challenges inherent in standard dark-siren analyses. 

Note that we neglect the impact of the source-frame mass. In other words, we are implicitly assuming that the detector distribution of source masses is uniform. Indeed, this is also the choice we made when generating \ac{gw} events; there could have been the possibility of introducing a bias for the cosmological inference. Therefore, the individual \ac{gw} events posterior samples of interest in our analysis are the luminosity distance and the sky position, namely RA and Dec. In this work, we deliberately did not use source-frame mass information in order to isolate the constraining power of the \ac{hi} redshift prior and avoid introducing dependence on a specific astrophysical mass model. Including the mass distribution, i.e., using spectral sirens, would provide an additional source of redshift information and could therefore improve cosmological constraints. However, it would also introduce new potential systematics~\citep{Mukherjee:2021rtw, Karathanasis:2022rtr, Agarwal:2024hld}.

\section{Results}\label{sec:results}

The two \ac{gw} catalogs (clustered and not), are analyzed either using or not the redshift information coming from the \ac{hi} maps. This comparison allows us to isolate the benefit of a correctly matched \ac{lss} prior and to test the bias introduced when the prior is inconsistent with the true source distribution.

\begin{figure*}
    \centering
    \includegraphics[width=1.35\columnwidth]{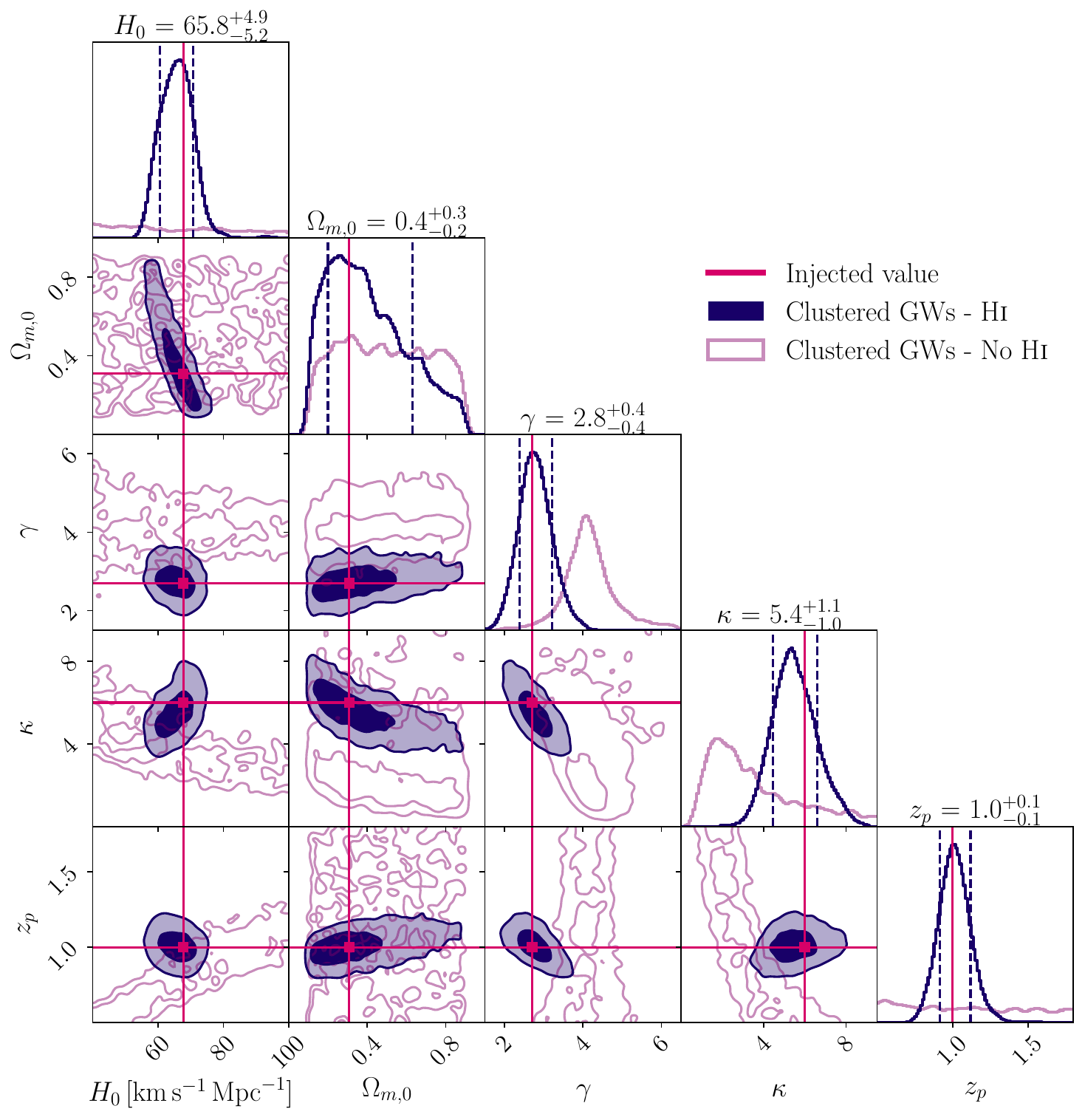}
    \caption{
    Corner plot showing the marginalized posterior distributions and the $1\sigma$ and $2\sigma$ 2D confidence regions for the cosmological and rate parameters $H_0$, $\Omega_{m,0}$, $\gamma$, $\kappa$, and $z_p$, inferred from clustered \ac{gw} events. The dark purple contours correspond to the analysis including \hi\ information (Clustered GWs - \hi), while the light purple contours show the case without \hi\ information (Clustered GWs - No \hi). The injected fiducial values are indicated by magenta lines. The panels on the diagonal display the one-dimensional marginalized posteriors, with quoted median values and 68\% credible intervals, highlighting the improvement in parameter constraints when \hi\ data are included.
    }
    \label{fig:h0_corner}
\end{figure*}

\subsection{Cosmology with Clustered \acp{gw}}

Our main result is that including \ac{hi} information substantially improves the constraint on the posterior of $H_0$ when the \ac{gw} events are clustered consistently with the underlying \ac{dm} field. For the Clustered GWs - \ac{hi} case, we obtain
\begin{equation}
H_0 = 65.8^{+4.9}_{-5.2}\ {\rm km\,s^{-1}\,Mpc^{-1}}
\end{equation}
at 68\% credibility, which represents a precision of $\sim 8\%$.

The result is shown in Fig.~\ref{fig:h0_corner}, which also reports the posterior on the other population parameters we considered in the analysis. The $H_0$ posterior is sharply peaked near the injected value, indicating that the \ac{hi} redshift prior carries sufficient information for the Hubble constant inference. By contrast, if the same clustered \ac{gw} catalog is analyzed without \ac{hi} information, the Hubble-constant posterior is much broader. In this case, the posterior is close to uninformative. The comparison demonstrates that adding \ac{hi} information transforms a weak constraint into a meaningful cosmological measurement, corresponding to an improvement of $\sim$90\%.
\\
The constraints on $\Omega_{m, 0}$ remain relatively weak even when \ac{hi} information is included. This reflects the limited sensitivity of the method to the shape of the distance–redshift relation beyond an overall scaling set by $H_0$.
\\
In terms of the other population parameters, we notice that in the case that \ac{hi} maps are used in the analyses, all injected values are properly recovered. Instead, if \ac{hi} maps are not included, and so the modeling of the \ac{los} prior does not take into account the redshift information coming from \ac{hi} maps, then we might obtain a systematic bias in the estimation of $k, z_p$ and $\gamma$. This systematic bias is introduced by the fact that the redshift distribution (number density) of \ac{gw} sources strongly deviates from a uniform in comoving volume distribution, in particular for lower redshifts.

\subsection{Cosmology with Isotropic \acp{gw}}

\begin{figure}
    \centering   
    \includegraphics[width=0.9\columnwidth]{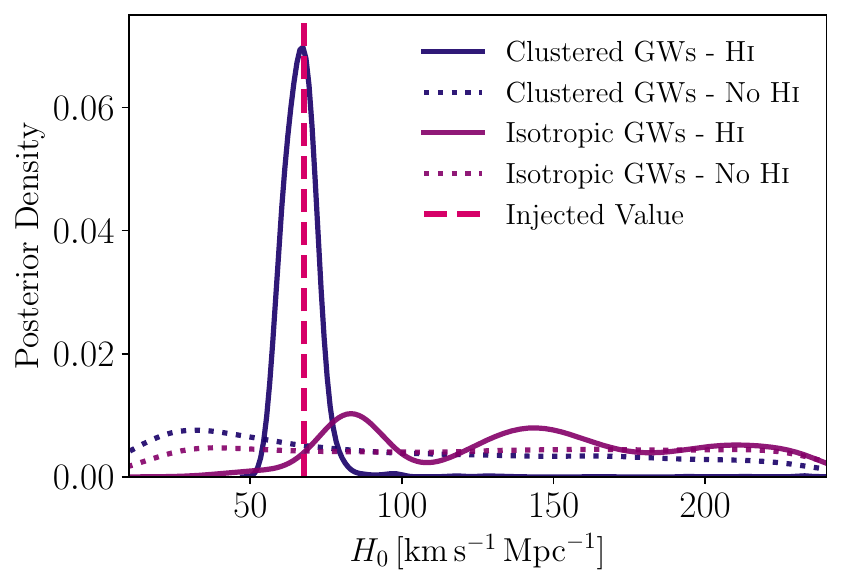}
    \caption{Posterior probability distributions for $H_0$ inferred from both clustered (Clustered GWs) and isotropically distributed (Isotropic GWs) \ac{gw} events under two scenarios: with \hi\ information included (\hi\,) and without \hi\ information (No \hi\,). The vertical line indicates the injected value of $H_0$.}
    \label{fig:headline_all}
\end{figure}

We also consider what would happen in the case that \ac{gw} events do not completely cluster as the \ac{dm} field and are instead isotropically distributed in the sky. In this case, as well, we study what happens when, at the inference level, we include or do not include the \ac{hi} prior.
We show the result of this test in Fig.~\ref{fig:headline_all}, where we compare with the clustered case. 
If the \ac{hi} clustering prior is still used in this case, the inferred Hubble constant becomes biased towards higher values.
If the isotropic catalog is analyzed without \ac{hi} information, the constraint remains weak.

These results show that the clustering prior is only beneficial when it accurately reflects the true source distribution. Otherwise, it can systematically pull the inferred cosmology away from the injected value. 
If \ac{gw}s are not clustered, and a \ac{hi} map is used for the inference, then it is likely that we will overestimate the value of $H_0$. This bias is introduced by the fact that, as \acp{gw} are uniform in the comoving volume, they are more likely to be placed at higher redshifts (higher $H_0$) as the \ac{hi} maps are more uniform. This process introduces a high $H_0$ bias.

This result highlights a key systematic risk of the method: the inferred cosmology is conditional on the assumed clustering model. If the true \ac{gw} population deviates from the assumed bias relation, the redshift prior becomes misspecified, leading to biased cosmological inference. This underscores the need to jointly model cosmology and source clustering.

This case gives us a warning: if in the Universe, due to astrophysical effects, the \ac{hi} field is not a good tracer for the \ac{gw} distribution, the \textit{radio-siren} method could give biased results. On sufficiently large scales, however, our main assumption should hold.

\begin{figure*}
    \centering
    \includegraphics[width=1.5\columnwidth]{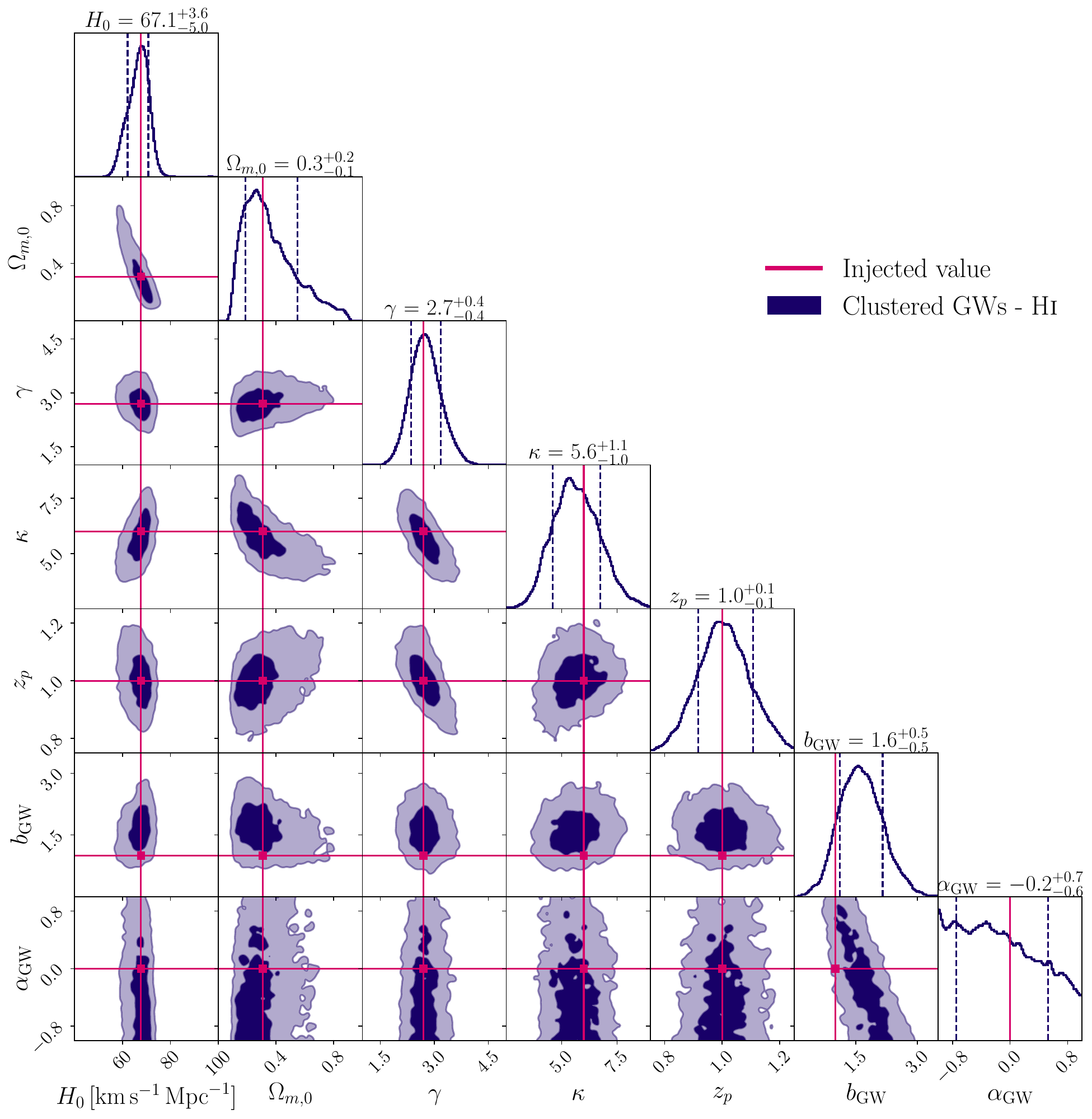}
    \caption{Results for the headline results with Clustered GWs - \hi\ including the two parameters entering the bias parametrization: $b_{\rm GW}$ and $\alpha_{\rm GW}$. The injected fiducial values are indicated with magenta lines.}
    \label{fig:bias_full}
\end{figure*}

\subsection{Validation test of the clustering of \ac{gw} sources}\label{sec:bias_gw}

\begin{figure}
    \centering
    \includegraphics[width=0.9\columnwidth]{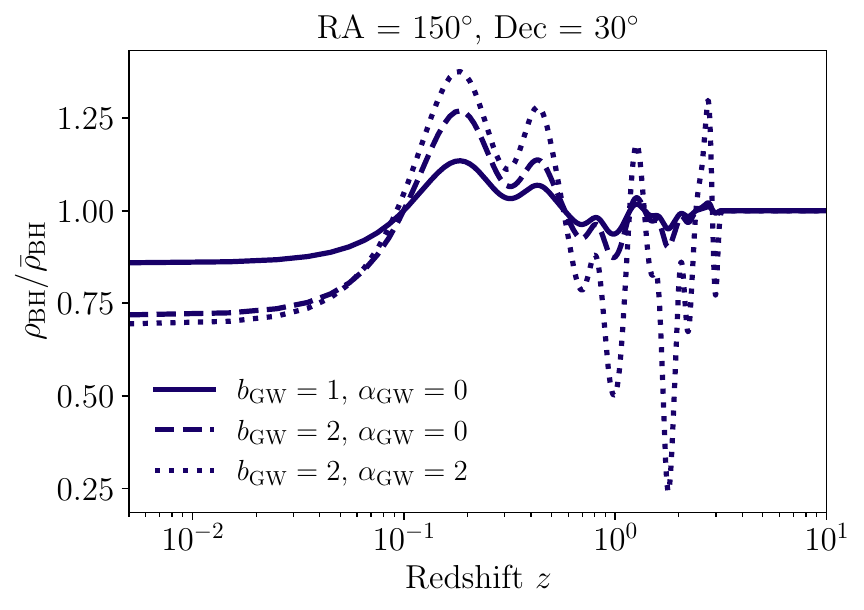}
    \caption{\Ac{los} evolution of the \ac{bh} mass density contrast, $\rho_{\rm BH}/\bar{\rho}_{\rm BH}$, as a function of redshift for a fixed sky direction (RA = $150^\circ$, Dec = $30^\circ$), under different assumptions for the \ac{gw} bias parameters. The baseline case ($b_{\rm GW} = 1$, $\alpha_{\rm GW} = 0$) corresponds to \ac{gw} sources tracing the underlying matter field without additional redshift dependence. Increasing the bias amplitude ($b_{\rm GW} = 2$) enhances the contrast of overdense and underdense regions, while introducing a redshift-dependent bias ($\alpha_{\rm GW} = 2$) further amplifies fluctuations at higher redshift.
}
    \label{fig:los_bias}
\end{figure}

An important consistency test of the pipeline is to simultaneously constrain the clustering properties of the \ac{gw} source population as in Eq.~\ref{eq:bias}. In particular, the bias parameters $(b_{\rm GW}, \alpha_{\rm GW}$) encode how \ac{bbh} mergers trace the underlying matter distribution.

In Fig.~\ref{fig:bias_full}, we explored this possibility by allowing these parameters to vary together with the other cosmological parameters. We find that we are able to recover the injected values within uncertainties, indicating that the method is self-consistent. Also, the posterior on $H_0$ is not affected by enlarging the number of population parameters we are doing inference on. 

To illustrate how the bias parameters affect the \ac{gw} distribution along the line of sight, we show, in Fig.~\ref{fig:los_bias}, how the \ac{bh} density contrast $\rho_{\rm BH}/\bar{\rho}_{\rm BH}$ evolves for different choices of $(b_{\rm GW}, \alpha_{\rm GW})$. Fixing a representative sky direction, we observe that increasing the bias amplitude enhances the contrast between overdense and underdense regions, effectively sharpening the redshift-dependent structure traced by the \ac{gw} population. 

When a redshift-dependent bias is introduced ($\alpha_{\rm GW} \neq 0$), the modulation becomes progressively stronger at higher redshift, leading to an amplification of fluctuations where the underlying matter field is otherwise more homogeneous. This behavior reflects the parametrization in Eq.~\ref{eq:bias}, where the bias enters as a multiplicative factor of the density contrast.

From the perspective of the \textit{radio-siren} framework, these effects directly translate into changes in the informativeness of the redshift prior. A stronger or evolving bias increases the contrast along the \ac{los}, potentially improving the localization of \ac{gw} events in redshift space. However, it also introduces additional modeling dependence: if the assumed bias does not match the true clustering properties of the source population, the inferred redshift prior may become misspecified, leading to biased cosmological constraints.

\subsection{Constraints on $H(z)$}

\begin{figure}
    \centering
    \includegraphics[width=0.9\columnwidth]{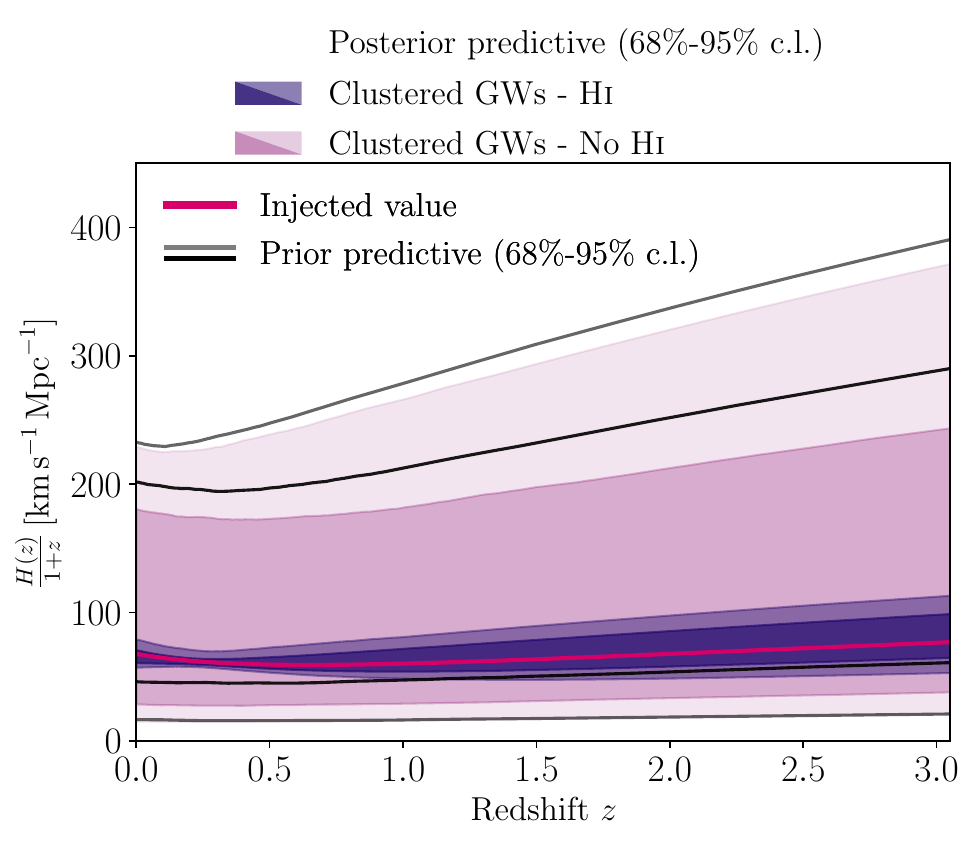}
    \caption{Posterior predictive constraints on $H(z)/(1+z)$ as a function of redshift $z$. The shaded bands show the $68\%$ (darker shade) and $95\%$ (lighter shade) credible intervals for Clustered GWs - \hi\, (in \textit{purple}) and for Clustered GWs - No \hi\, (in \textit{pink}). The solid magenta line indicates the injected cosmological model. For comparison, the black lines represent the prior predictive distribution ($68\%$--$95\%$ credible intervals).}
    \label{fig:h_of_z}
\end{figure}

Although the main focus of this work is the Hubble constant, the same posterior samples can be propagated to reconstruct the late-time expansion history over the redshift range covered by the \ac{hi} maps. Assuming a flat $\Lambda$CDM background, the expansion rate is obtained from the inferred cosmological parameters as:
\begin{equation}\label{eq:hofz}
H(z)=H_0\sqrt{\Omega_{\rm m,0}(1+z)^3 + 1-\Omega_{\rm m,0}}
\end{equation}
Fig.~\ref{fig:h_of_z} shows the posterior predictive reconstruction of $H(z)/(1+z)$ as a function of redshift, comparing the Clustered GWs - \hi\ and the Clustered GWs - No \hi\ analyses. The shaded regions correspond to the 68\% (darker shade) and 95\% (lighter shade) credible intervals, while the prior predictive distribution is also shown for reference (black lines).

The inclusion of \ac{hi} information leads to a substantial tightening of the reconstructed expansion history, particularly at low redshift ($z \lesssim 0.5$), where the density contrast of the \ac{hi} field is highest and the redshift prior is most informative. In this regime, the posterior predictive band for the Clustered GWs - \hi\ case closely traces the injected cosmology, with a significant reduction in uncertainty relative to the prior and to the Clustered GWs - No \hi\ . This reflects the improved constraint on $H_0$, which sets the overall normalization of the expansion rate.

At intermediate redshifts ($0.5 \lesssim z \lesssim 2$), the reconstruction remains consistent with the fiducial model, although the uncertainty gradually increases. Toward the highest redshifts probed by the \ac{hi} maps ($z \sim 3$), the posterior predictive distribution broadens and approaches the prior envelope, indicating that the data carry limited information on the detailed shape of $H(z)$ in this regime. This is consistent with the relatively weak constraints on $\Omega_{m,0}$ observed in Fig.~\ref{fig:h0_corner}, and highlights that the current setup primarily constrains the normalization of the expansion history rather than its evolution.

Overall, these results show that the \textit{radio-siren} method not only constrains $H_0$, but also enables a consistent reconstruction of the late-time expansion history over the redshift range $z \in [0, \sim 3]$. While the sensitivity to the detailed redshift evolution remains limited in the present configuration, future analyses with larger \ac{gw} catalogs and more realistic \ac{hi} modeling may allow for competitive constraints on $H(z)$ beyond a simple normalization measurement.

\section{Discussion and Conclusions}\label{sec:disc_and_concl}

We have presented a proof-of-concept study of \textit{radio sirens}, a joint cosmological framework that combines \ac{bbh} standard sirens with \ac{hi} \ac{im}. The central idea is that, in the absence of \ac{em} counterparts, the \ac{lss} traced by \ac{hi} can be used as a redshift prior for dark sirens. Using simulated \ac{et} \ac{bbh} events and mock SKA-Mid-like \ac{hi} maps, we find that this additional information dramatically improves cosmological inference when the source population is clustered consistently with the underlying \ac{hi} field. 

Our fiducial clustered analysis yields a constraint on $H_0$ at the $\sim 8\%$ level, improving by $\sim 90\%$ with respect to the corresponding Clustered \acp{gw} - No \hi\ case, which is largely uninformative. This demonstrates that tomographic 21\,cm maps can be used for dark siren cosmology: a physically motivated redshift prior that exploits the three-dimensional structure of the Universe. The method is especially promising in the \ac{et} era, when the number of \ac{bbh} detections will be sufficiently large to enable population-level inference out to cosmological distances~\citep{Branchesi:2023mws, ET:2025xjr}.

At the same time, our isotropic-source tests highlight an important caveat. If the \ac{gw} source population does not follow the matter distribution in the way assumed by the prior, then the \ac{hi}-informed analysis can introduce substantial bias. This means that future applications of the \textit{radio-siren} analysis must rely on realistic astrophysical models for \ac{bbh} environments and clustering.

Additionally, several simplifying assumptions made here should be relaxed in future work. To begin with, we have assumed that the \ac{hi} field traces the underlying matter density without additional bias or stochasticity. At the single \ac{gw}-event analysis, we relied on the Fisher matrix approximation, even though enhanced through the inclusion of a simplified version of priors~\citep{Dupletsa:2024gfl}. 

Another important extension of the present analysis would concern the inclusion of the \ac{bbh} mass spectrum. A realistic implementation should jointly model the mass spectrum, the \ac{bbh} merger-rate evolution, and the clustering relation between \ac{gw} sources and the \ac{hi} density field.

Finally, our treatment neglects observational complications in \ac{im}, such as beam effects, foreground removal, calibration systematics, and the detailed response of SKA-Mid. A more realistic end-to-end analysis should also include these effects.

Despite these limitations, the present study establishes the basic viability of the \textit{radio-siren} concept. With the large \ac{bbh} catalogs expected from \ac{3g} detectors and the tomographic reach of future 21\,cm surveys, the joint \ac{et}-SKA approach could become a powerful and independent route to precision cosmology.

An additional advantage of the \textit{radio-siren} framework concerns the treatment of selection effects and completeness. In traditional galaxy-catalog dark-siren analyses, the redshift prior relies on discrete galaxy samples, often combined from different surveys and inherently flux-limited. This requires careful modeling of catalog incompleteness, typically through assumptions on the galaxy luminosity function and the evaluation of the number of missing galaxies as a function of redshift. Such corrections introduce both statistical uncertainty and potential systematic biases, especially at higher redshifts where incompleteness becomes severe.

In contrast, the use of \ac{hi} intensity mapping significantly simplifies this aspect. Since the observable is a finely redshift-spaced tracer of the underlying matter density field, rather than a catalog of individually detected sources, the need for explicit completeness corrections is largely removed. At leading order, the redshift prior can be constructed assuming a distribution uniform in comoving volume, whenever the information from \ac{hi} density contrast is not available. This advantage becomes even more important at high redshift. While galaxy-catalog dark-siren analyses become increasingly limited by catalog incompleteness and heterogeneous survey selection functions, \ac{hi} intensity mapping provides a tomographic tracer of the \ac{lss} over large cosmological volumes. This is particularly relevant for next-generation detectors, where the combination of large volumes and high redshift coverage would otherwise make completeness corrections increasingly challenging and model-dependent.

Overall, \textit{radio sirens} should be viewed as an additional source of information. Galaxy catalogs provide discrete host-galaxy information, but become increasingly limited by completeness and survey-selection effects at high redshift. The \ac{bbh} mass spectrum can provide additional statistical redshift information, but requires careful modeling of astrophysical evolution and of possible redshift dependence in the source population. \ac{hi} intensity mapping supplies a spectroscopic tracer of the \ac{lss} over large cosmological volumes that is not completeness-limited, as the full \ac{hi} mass function is sampled in the measured temperature brightness of each pixel. 
This makes it particularly well matched to the high-redshift reach of third-generation \ac{gw} detectors. Combining these ingredients in a joint framework will be essential to exploit the full potential of \ac{bbh} standard sirens, while controlling the associated astrophysical, observational, and population-modeling systematics.

\section*{Acknowledgements}

The authors acknowledge support from the \href{https://www.sgws-community.eu}{Sardinian GW Science (SGWS) Community} through a joint program between the University of Cagliari and the Gran Sasso Science Institute (GSSI) in the framework of activities supporting Sardinia's candidacy to host the \href{https://www.einstein-telescope.it/srd/home-sar/}{Einstein Telescope}.
% ERC Simone
S.~M. is supported by the ERC grant GravitySirens  101163912. Funded by the European Union. Views and opinions expressed are, however, those of the author(s) only and do not necessarily reflect those of the European Union or the European Research Council Executive Agency. Neither the European Union nor the granting authority can be held responsible for them.
M.~S. is supported by the French government through the France 2030 investment plan managed by the National Research Agency (ANR), as part of the Initiative of Excellence Université Côte d’Azur under reference number ANR- 15-IDEX-01.
D.~N-L thanks the Spanish MCIN/AEI/10.13039/501100011033 under the Grants No.~PID2020-113701GB-I00 and PID2023-146517NB-I00, some of which include ERDF funds from the European Union, and by the MICINN with funding from the European Union NextGenerationEU (PRTR-C17.I1) and by the Generalitat de Catalunya. IFAE is partially funded by the CERCA program of the Generalitat de Catalunya. D.~N-L has also received funding from the European Union’s Horizon Europe research and innovation programme under the Marie Skłodowska-Curie grant agreement No.~10181337.
CC acknowledges the Italian Ministry of University and Research (MUR) PRIN 2022 “EXSKALIBUR – Euclid-CrossSKA: Likelihood Inference Building for Universe’s Research”, Grant No. 20222BBYB9, CUP C53D2300083 0006, from the European Union – Next Generation EU. MC is partially supported by the 2025/26 ``Research and Education'' grant from Fondazione CRT. MC's institute (OAVdA) is managed by the Fondazione Cl\'ement Fillietroz, which is supported by the Regional Government of the Aosta Valley, the Town Municipality of Nus and the ``Unit\'e des Communes vald\^otaines Mont-\'Emilius''.

%%%%%%%%%%%%%%%%%%%%%%%%%%%%%%%%%%%%%%%%%%%%%%%%%%
\section*{Data Availability}
\textit{The data products, samples of the \ac{gw} events, the injection set and the \ac{hi} maps, are available on Zenodo (upon publication of this paper). Tutorials on their generation and usage are included. \href{https://github.com/icarogw-developers/icarogw/tree/HI}{Link} to \ac{hi} branch of icarogw.}

\bibliographystyle{mnras}
\bibliography{references} % if your bibtex file is called example.bib
%%%%%%%%%%%%%%%%%%%%%%%%%%%%%%%%%%%%%%%%%%%%%%%%%%

%%%%%%%%%%%%%%%%% APPENDICES %%%%%%%%%%%%%%%%%%%%%

\appendix

\section{Additional validation tests}\label{app:validation_tests}
We performed two additional validation tests for both the Isotropic GWs - \hi\ and Clustered GWs - \hi\ analyses discussed in the main text.

\begin{figure}
    \centering
    \includegraphics[width=0.9\columnwidth]{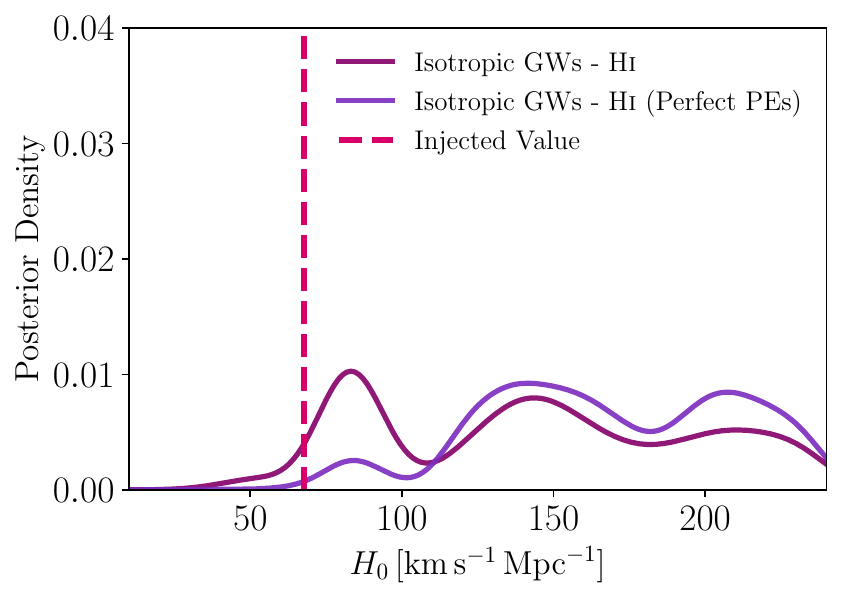}
    \caption{Posterior probability distributions for $H_0$ inferred from isotropically distributed (Isotropic GWs) \ac{gw} events under two with \hi\ information included (\hi\ ) at the analysis stage. We show the results with perfectly measured \acp{gw} and including \ac{pe} errors. The vertical line indicates the injected value of $H_0$.}
    \label{fig:gw_isotropic_perfectPE}
\end{figure}

First, we revisited the Isotropic GWs - \hi\ case by comparing the analysis based on Fisher-matrix parameter-estimation uncertainties with an idealized setup in which the \ac{gw} events are assumed to be perfectly measured. The result is shown in Fig.~\ref{fig:gw_isotropic_perfectPE}. The posterior remains biased when the \ac{hi} prior is included even in the perfect-\ac{pe} limit, demonstrating that the dominant source of the shift is not measurement noise, but rather the mismatch between the assumed clustering prior and the true source distribution. In other words, if the \ac{bbh} population does not trace the matter field while the inference model assumes that it does, the redshift prior is misspecified and the cosmological posterior is shifted accordingly.

Interestingly, measurement errors on the \ac{gw} side help the $H_0$ posterior to shift towards the injected value. Once the single-event likelihoods are broadened in luminosity distance and sky position, the posterior support is distributed over a wider region of the redshift prior, which slightly reduces the tendency of the analysis to concentrate probability at the highest redshifts allowed by the nearly homogeneous \ac{hi} maps. This explains why the biased posterior obtained with Fisher uncertainties is somewhat closer to the injected value than in the perfect-\ac{pe} case.

\begin{figure}
    \centering
    \includegraphics[width=0.9\columnwidth]{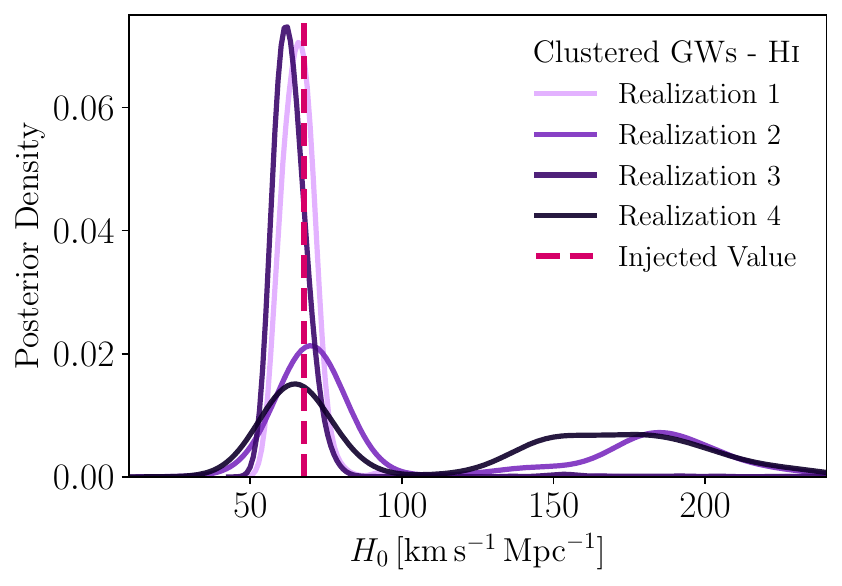}
    \caption{Posterior probability distributions for $H_0$ inferred from clustered \ac{gw} events with \hi\ information included, under different \ac{gw} events distribution realizations. The vertical line indicates the injected value of $H_0$.}
    \label{fig:h0_random_seed}
\end{figure}

Second, we tested the robustness of the headline Clustered GWs - \hi\ result against fluctuations in the source catalog. To do so, we generated several independent realizations of the clustered \ac{gw} population. Figure~\ref{fig:h0_random_seed} shows that the recovered $H_0$ posterior is stable across realizations: all runs exhibit a dominant peak close to the injected value, with only moderate realization-to-realization scatter. A subset of the realizations additionally shows a secondary peak extending toward higher $H_0$, indicating that finite-sample fluctuations can still produce some multi-modality at the catalog size considered here. Overall, these tests support the conclusion that the main qualitative behavior discussed in Sec.~\ref{sec:results} is not driven by a single lucky or pathological realization.

\section{Simulation of the BBH events}\label{app:gw_simulation}
We generate a synthetic population of \ac{bbh} mergers assuming the merger-rate model introduced in Sec.~\ref{sec:gw_sim}. The fiducial cosmology is fixed to the values reported in Tab.~\ref{tab:priors_table}, namely $H_0=67.7\,{\rm km\,s^{-1}\,Mpc^{-1}}$ and $\Omega_{m,0}=0.308$, while the redshift evolution of the source-frame rate is described by the modified Madau--Dickinson parameters $\gamma=2.7$, $\kappa=6$, and $z_p=1$. These values were chosen to place a substantial fraction of the detectable population within the redshift interval covered by the \ac{hi} maps, rather than maximizing the total number of \ac{et} detections. For this reason, the catalog contains fewer events than would be expected in a fully realistic one-year \ac{et} forecast ~\citep[e.g.][]{Branchesi:2023mws, ET:2025xjr}, but it is better matched to the science target of this study.

We simulate $10^4$ \ac{bbh} mergers, corresponding to one year of observations for the adopted rate normalization. Sky position and distance are assigned using the mock \ac{lss} information. In the clustered catalog, events are distributed according to the density-modulated merger-rate field of Eqs.~\ref{eq:madau} and~\ref{eq:bias}, so that \ac{bbh} mergers trace the same underlying \ac{dm} structures. In the isotropic catalog, by contrast, the clustering term is switched off, and the sources are distributed uniformly in comoving volume. The two catalogs, therefore, share the same detector setup and selection effects, but differ only in whether they explicitly follow the \ac{dm} distribution. All the other binary parameters are sampled from isotropic or uniform distributions over their physical ranges as detailed in Tab.~\ref{tab:gw_ranges}.

\begin{table}
  \caption[]{List of the prior distribution for the parameters describing a \ac{gw} event used to generate the starting datasets. For a definition of the listed parameters, see Tab.~4 in~\cite{Dupletsa:2024gfl}. Note that here we provide information on the redshift distribution, whereas at the \ac{gw} \ac{pe} level we use the luminosity distance. The corresponding luminosity distance values are obtained from the redshift assuming our injected cosmology as in Tab.~\ref{tab:priors_table}.}
  \label{tab:gw_ranges}
  \begin{tabular}{l | c c c}
  \toprule
    \textbf{parameter} &\textbf{units} &\textbf{prior} &\textbf{prior range} \\
    \midrule
    $m_1$ &$M_{\odot}$ &Uniform &$[5, 300]$\\
    $m_2$ &$M_{\odot}$ &Uniform &$[5, 300]$\\
    $z$ &[Mpc] &\makecell[c]{Uniform in source-frame \\ + Madau-Dickinson (+\ac{hi})} &$[0, \sim 3]$\\
    $\theta_{JN}$ &[rad] &Sine &$[0, \pi]$\\
    Dec &[rad] &Cosine &$[-\frac{\pi}{2}, +\frac{\pi}{2}]$\\
    RA &[rad] &Uniform &$[0, 2\pi]$\\
    $\phi$ &[rad] &Uniform &$[0, 2\pi]$\\
    $\Psi$ &[rad] &Uniform &$[0, \pi]$\\
    $t_c$ &[s] &Delta &$0.$\\
    $a_1$ &- &Uniform &$[0, 0.99]$\\
    $a_2$ &- &Uniform &$[0, 0.99]$\\
    \texttt{tilt}$_1$ &[rad] &Sine &$[0, \pi]$\\
    \texttt{tilt}$_2$ &[rad] &Sine &$[0, \pi]$\\
    \texttt{phi}$_{12}$ &[rad] &Uniform &$[0, 2\pi]$\\
    \texttt{phi}$_{JL}$ &[rad] &Uniform &$[0, 2\pi]$\\
    \bottomrule
  \end{tabular}
\end{table}

The generated events are passed through \texttt{GWFish} using \ac{et} in its triangular configuration and the full cryogenic sensitivity curve~\cite{Branchesi:2023mws}. Even though we retain only the events with a \ac{snr} greater than 150, an unconstrained multivariate Gaussian approximation can place posterior support outside the physical parameter domain, for example, at negative luminosity distance or outside the allowed angular ranges. To avoid this, we post-process the Fisher covariance by drawing samples from a truncated multivariate Gaussian, following the procedure described in~\cite{Dupletsa:2024gfl}. This step preserves the local Fisher information while enforcing the physical priors required by the hierarchical analysis. We apply uniform cuts in the ranges detailed in Tab.~\ref{tab:gw_ranges}, except for the masses and distance, where we impose a positivity requirement only.

\begin{figure}
    \centering
    \includegraphics[width=0.9\columnwidth]{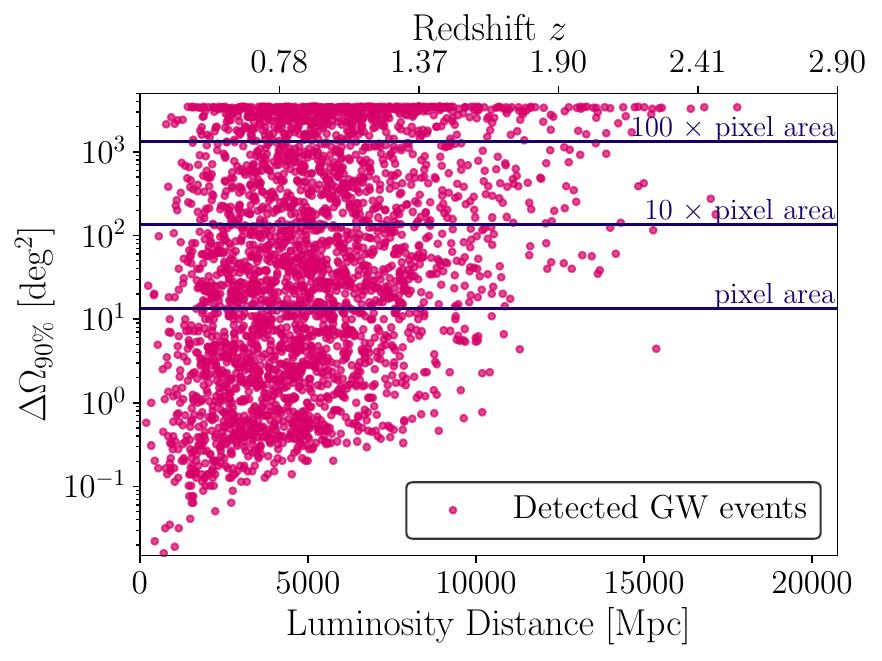}
    \caption{Luminosity distance versus sky localization area for the $\sim 3100$ detected \ac{gw} events in the clustered \acp{gw} case. Each point represents a detected event, plotted as a function of luminosity distance (bottom axis) and corresponding redshift (top axis). The vertical axis shows the 90\% credible sky localization area, $\Delta \Omega_{90\%}$, in deg$^2$. Horizontal lines indicate the pixel area of the map ($\sim 13$\,deg$^2$) and multiples thereof (10\,$\times$ and 100\,$\times$) for reference. The plot illustrates the general trend of increasing localization uncertainty with distance (and redshift). 
    }
    \label{fig:sky_loc_distance_scatter}
\end{figure}

The posterior samples produced in this way are then used as input to \texttt{icarogw}. For each event, the relevant information entering the population analysis is the luminosity distance and sky position (RA and Dec) samples. Fig.~\ref{fig:sky_loc_distance_scatter} summarizes the localization properties of the detected sample of \ac{gw} events by showing the luminosity distance versus sky localization area. Here we report the headline Clustered GWs - \hi\ results which contain $\sim 3100$ detected \ac{gw} events (the events simulated for the Isotropic GWs - \hi\ scenario present analogous distance-sky location distribution). Each point represents a detected event, plotted as a function of luminosity distance and corresponding redshift. The vertical axis shows the 90\% credible sky localization area, $\Delta \Omega_{90\%}$, in deg$^2$. The sky localization has been computed from the Fisher samples with prior cuts, with nside progressively increased from 64 to 2048 for increasingly well-localized \ac{gw} events. Horizontal lines indicate the pixel area of the map ($\sim 13$\,deg$^2$) and multiples thereof (10\,$\times$ and 100\,$\times$) for reference. The plot illustrates the general trend of increasing localization uncertainty with distance (and redshift).

\section{Breaking down hierarchical Bayesian inference and its numerical implementation in \texttt{icarogw}}\label{app:icaro}

The hierarchical analysis is performed with \texttt{icarogw}, which evaluates the population likelihood by Monte Carlo reweighting of both posterior samples and injections~\cite{Mastrogiovanni:2023zbw}. Since the single-event likelihood $\mathcal{L}_{\rm GW}(x_i\mid\theta,\Lambda)$ is not directly available, the event term in Eq.~\eqref{eq:hierarchical_lkh} is estimated from posterior samples by undoing the prior used in the parameter-estimation stage. Denoting by $N_{s,i}$ the number of posterior samples for event $i$, one obtains: 
\begin{equation}
    \label{eq:numerical_hierachical_lkh}
    \begin{aligned}
    \int \mathcal{L}_{\rm GW}\left(x_i|\theta, \Lambda \right)\frac{\dd N_{\rm CBC}}{\dd t \dd\theta}(\Lambda)\dd\theta &\sim \frac{1}{N_{s,i}} \sum_{j=1}^{N_{s,i}} \frac{1}{\pi(\theta_{i,j}|\Lambda)}\frac{\dd N_{\rm CBC}}{\dd \theta \dd t}(\Lambda)|_{i,j} \\
    &\equiv \frac{1}{N_{s,i}} \sum_{j=1}^{N_{s,i}} w_{i,j}
    \end{aligned}
\end{equation}
where $\pi(\theta_{i,j}|\Lambda)$ is the prior used to obtain the posterior samples.

Selection effects are treated in an analogous way using the injection set. If $N_{\rm gen}$ is the total number of simulated injections and $N_{\rm det}$ the number of injections that pass the detection threshold, the expected number of detections can be estimated as:
\begin{equation}
    \label{eq:numerical:pdet}
    \begin{aligned}
        N_{\rm exp} &= T_{\rm obs} \int p_{\rm det} (\theta) \frac{\dd N_{\rm CBC}}{\dd t\dd\theta}(\Lambda) \dd \theta\\
        &\sim \frac{T_{\rm obs}}{N_{\rm gen}} \sum_{j=1}^{N_{\rm det}} \frac{1}{\pi_{\rm inj}(\theta_j)} \frac{\dd N_{\rm CBC}}{\dd t \dd\theta}|j\\
        &\equiv \frac{T_{\rm obs}}{N_{\rm gen}}  \sum_{j=1}^{N_{\rm det}} s_j
    \end{aligned}
\end{equation}
where $\pi_{\rm inj}(\theta_j)$ is the prior used to generate the \textit{injections}.

The final numerical expression for Eq.~\eqref{eq:hierarchical_lkh}, as computed in \texttt{icarogw}, is given by:
\begin{equation}
    \label{eq:numerical_lkh_all}
    \ln \mathcal{L}(\{x\}|\Lambda) \sim -\frac{T_{\rm obs}}{N_{\rm gen}}  \sum_{j=1}^{N_{\rm det}} s_j +  \sum_{i=1}^{N_{\rm obs}}\ln \left[ \frac{T_{\rm obs}}{N_{s,i}} \sum_{j=1}^{N_{s,i}} w_{i,j}\right]
\end{equation}

A useful way to assess the numerical robustness of this Monte-Carlo estimate is through effective-sample-size diagnostics. For the reweighted posterior samples of a single event $i$ we define:
\begin{equation}
N_{\rm eff,PE}^{(i)} = \frac{\left(\sum_j w_{i,j}\right)^2}{\sum_j w_{i,j}^2},
\end{equation}
which measures how many posterior samples contribute effectively after reweighting~\cite{Talbot:2023pex}. Likewise, for the injection term, we define:
\begin{equation}
N_{\mathrm{eff,inj}} =
\frac{\left(\sum_{j}^{N_{\mathrm{det}}} s_j \right)^2}
{\sum_{j}^{N_{\mathrm{det}}} s_j^2 - \frac{1}{N_{\mathrm{gen}}}\left(\sum_{j}^{N_{\mathrm{det}}} s_j \right)^2}
.
\end{equation}
which quantifies the effective support of the injection set in the region of hyper-parameter space favored by the analysis. 

\begin{figure}
    \centering
    \includegraphics[width=0.9\columnwidth]{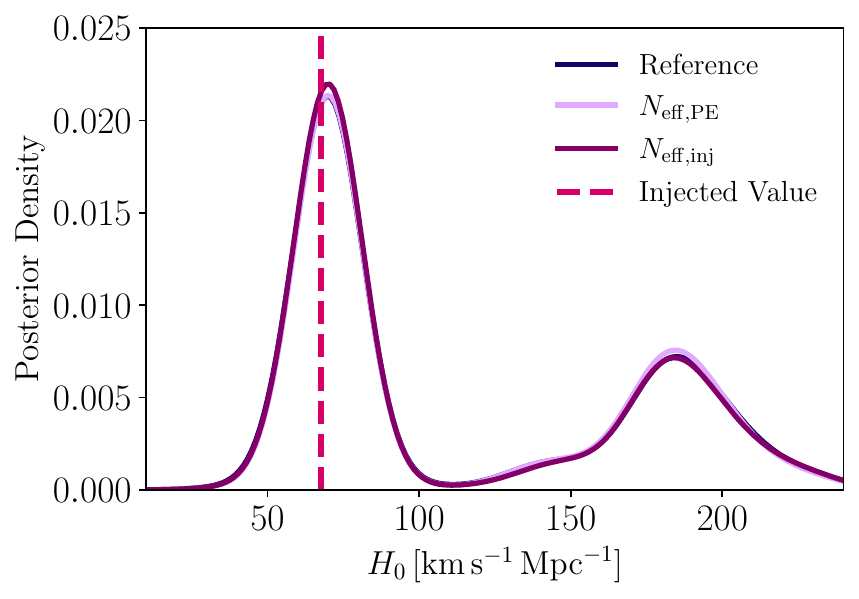}
    \caption{Posterior probability distributions for $H_0$ inferred from clustered \ac{gw} events with \hi\ information included (\hi\ ), under two different numerical stability checks: $N_{\rm eff, PE}$ and $N_{\rm eff, inj}$. The vertical line indicates the injected (true) value of $H_0$.}
    \label{fig:h0_checks}
\end{figure}

In Fig.~\ref{fig:h0_checks} we compare the reference Clustered GWs - \hi\ analysis with runs in which we explicitly monitor the impact of these two diagnostics. The posterior on $H_0$ remains consistent across the different choices, indicating that the headline result is not driven by a numerical pathology of the reweighting procedure.

%%%%%%%%%%%%%%%%%%%%%%%%%%%%%%%%%%%%%%%%%%%%%%%%%%
\bsp	% typesetting comment
\label{lastpage}
\end{document}